\documentclass[conference,a4paper]{IEEEtran}
\IEEEoverridecommandlockouts
\usepackage{cite}
\usepackage{amsmath,amssymb,amsfonts}
\usepackage[caption=false,font=footnotesize,labelfont=sf,textfont=sf]{subfig}
\usepackage[linesnumbered,ruled]{algorithm2e}
\usepackage{algpseudocode}
\usepackage{multirow}
\usepackage{graphicx}
\usepackage{textcomp}
\usepackage{xcolor}
\usepackage[colorlinks=true,linkcolor=black,citecolor=black,filecolor=black,urlcolor=black]{hyperref}
\usepackage{flushend}

\def\BibTeX{{\rm B\kern-.05em{\sc i\kern-.025em b}\kern-.08em
    T\kern-.1667em\lower.7ex\hbox{E}\kern-.125emX}}

\def\todo[#1]{\textit{\textcolor{blue}{[TODO: #1]}}}
\begin{document}

\title{Explainability of CNN Based Classification Models for Acoustic Signal%*\\
%{\footnotesize \textsuperscript{*}Note: Sub-titles are not captured in Xplore and
%should not be used
%}
% \thanks{This work was funded by EPSRC, UK, grant reference EP/V007335/1.}
}

% \author{\IEEEauthorblockN{Anonymous Authors}} 

\author{
\IEEEauthorblockN{
Zubair Faruqui\IEEEauthorrefmark{1}, Mackenzie S. McIntire\IEEEauthorrefmark{2}, Rahul Dubey\IEEEauthorrefmark{1}, Jay McEntee\IEEEauthorrefmark{3}
}
\IEEEauthorblockA{\IEEEauthorrefmark{1}Department of Computer Science, Missouri State University, Springfield, MO 65897 USA}
\IEEEauthorblockA{\IEEEauthorrefmark{2}Department of Biology, Missouri State University, Springfield, MO 65897 USA}
\IEEEauthorblockA{\IEEEauthorrefmark{3}Biodiversity Research Institute, Portland, ME 04103 USA}
% \IEEEauthorblockA{\{rahuldubey, bioauthor1, bioauthor2, bioauthor3\}@missouristate.edu}

}

\maketitle

\begin{abstract}

% Explainable AI (XAI) has gained significant attention due to its ability to interpret predictions of complex deep learning models. One domain benefiting from deep models is bioacoustics, where they have been effectively applied to classify birdsongs. However, limited research exists on the explainability of bioacoustic models. This paper analyzes the songs of Bewick’s wren, a bird species with recordings collected from the Southwest United States. The audio data was converted into spectrogram images and used to train a deep Convolutional Neural Network (CNN) for binary classification, achieving an accuracy of 94.5\%. To interpret the model’s predictions, both model-agnostic and model-specific XAI techniques, including LIME, SHAP, DeepLift, and Grad-CAM, were applied. A detailed comparison of the resulting explanations is presented in the results section. Findings indicate that Grad-CAM provided consistent and reliable explanations by highlighting the key regions of the spectrogram images that distinguish the two classes, whereas LIME's explanations were often inconsistent and deemed inaccurate by domain experts. This approach can be applied to other bioacoustics problems, including species identification, habitat monitoring, and detecting animal vocalizations, where interpretability is crucial for ecological and conservation studies. This work highlights the importance of selecting effective XAI methods for improving trust and interpretability in bioacoustic classification tasks.

Explainable Artificial Intelligence (XAI) has emerged as a critical tool for interpreting the predictions of complex deep learning models. While XAI has been increasingly applied in various domains within acoustics, its use in bioacoustics, which involves analyzing audio signals from living organisms, remains relatively underexplored. In this paper, we investigate the vocalizations of a bird species with strong geographic variation throughout its range in North America. Audio recordings were converted into spectrogram images and used to train a deep Convolutional Neural Network (CNN) for classification, achieving an accuracy of 94.8\%. To interpret the model’s predictions, we applied both model-agnostic (LIME, SHAP) and model-specific (DeepLIFT, Grad-CAM) XAI techniques. These techniques produced different but complementary explanations, and when their explanations were considered together, they provided more complete and interpretable insights into the model’s decision-making. This work highlights the importance of using a combination of XAI techniques to improve trust and interoperability, not only in broader acoustics signal analysis but also argues for broader applicability in different domain specific tasks. 

% and classification tasks, with potential applications in species identification, habitat monitoring, and wildlife conservation.

\end{abstract}


\section{Introduction}

Acoustic research, encompassing the study of sound signals produced by a wide range of sources, including humans, animals, and natural environments, offers critical insights into communication, behavior, and environmental monitoring. Acoustic signals are mechanical waves that carry information through variations in pressure over time, and their analysis is fundamental to a wide range of applications, including human speech analysis~\cite{kent2008acoustic}, bioacoustics monitoring~\cite{lee2021biosignal}, structural health diagnostics~\cite{zhang2022vibration}, and environmental sensing~\cite{kroos2019generalisation}. These approaches enable the study of communication, behavior, and physical processes across both biological and engineered systems. A significant area of focus involves analyzing biological signals, such as vocalizations from mammals, birds, and marine species, which serve important functions like mate attraction, territorial defense, and social interaction~\cite{Kroodsma1991}. While Deep Learning (DL) models have shown promise in classifying and detecting acoustic signals across these domains~\cite{zaman2023survey}, the lack of interpretability in their decision-making processes poses challenges for understanding the underlying features driving predictions~\cite{das2024exploring, akman2024audio}. This limitation underscores the growing need for explainable AI (XAI) techniques to enhance the transparency, reliability, and scientific value of machine learning models in bioacoustics signal analysis.

% \textcolor{red}{Talk about your experimental setup and provide brief results.  DONE}

Motivated by the challenge of interpretability and explainability of deep models in bioacoustic research, this paper presents a novel ensemble XAI approach to explain the predictions made by a  deep Convolutional Neural Network (CNN) trained in a supervised learning manner. For this purpose, we collected data of Bewick's wren (\textit{Thryomanes bewickii}), a small songbird native to the American Southwest that displays large-scale variation in song throughout its geographic range \cite{kroodsma1974song}. This work focuses on two major regional variants, referred here as \textit{Eastern} and \textit{Mexican} variants. These variants contain considerable within-group variation (sub-groups) as well. Each individual male produces many different songs, regardless of whether they sing Eastern or Mexican-type songs. This study aims to leverage XAI to characterize the complex variation between these groups. After performing sound recording of Bewick's wrens in Arizona and New Mexico, 2,660  4-second segments containing song samples from Bewick's wrens were converted to spectrogram images by Short-Time Fourier Transform on the audio signals\cite{seewave2008}. 

% A deep CNN was hypertuned with 5 convolutional layers with ReLU activation and max-pooling, followed by 3 fully connected layers to train on these image classes. We found a maximum prediction accuracy of 94.8\% on the dataset.

To evaluate the suitability and effectiveness of different XAI techniques in the bioacoustic domain, both model-agnostic (LIME\cite{ribeiro2016Lime}, SHAP\cite{lundberg2017shap}) and model-specific (Grad-CAM\cite{selvaraju2017gradcam}, DeepLIFT\cite{shrikumar2017deeplift}) explanation methods were applied to interpret model predictions. The model-specific techniques provided consistent and biologically meaningful explanations by highlighting the most relevant song signals. In contrast, model-agnostic methods often produced inconsistent and less interpretable results. To further enhance interpretability, ensemble saliency maps were constructed by combining Grad-CAM and DeepLIFT explanations (heatmaps), capturing the complementary regions emphasized by both techniques. The ensemble of Grad-CAM and DeepLIFT presented in the paper can be extended to other domain-specific task as well, showcasing the applicability and scalability of the proposed approach. 

Finally, to further investigate sub-grouping within given \textit{Eastern} and \textit{Mexican} variants, analysis using t-SNE\cite{tsne} and principal component analysis (PCA)\cite{pca} were conducted, which revealed the presence of distinct clusters within individual classes/variants.
% indicating that the model had captured subtle intra-class variations. 
These results demonstrate the value of XAI in uncovering fine-grained acoustic patterns and suggest its potential for broader applications in ecological monitoring, species classification, and hypothesis generation in bioacoustic research.

% This paper presents a deep learning-based approach for the classification of bird song recordings collected from various geographical locations, and applies XAI techniques to interpret the model’s predictions and to find hidden patterns in biological data. We explored four different XAI techniques: Local Interpretable Model-agnostic Explanations (LIME)~\cite{ribeiro2016Lime} and Shapley Additive Explanations (SHAP)~\cite{lundberg2017shap}, Gradient-weighted Class Activation Mapping (Grad-CAM)~\cite{selvaraju2017gradcam} and Deep Learning Important FeaTures (DeepLIFT)~\cite{shrikumar2017deeplift}. These techniques were selected to provide both local and global interpretability and to capture spatially relevant regions within the input spectrograms that influence the model's classification decisions. 

% Acoustics researchers have used deep CNN models to identify vocalizations of different birds' songs/signals, like bats and other animals. Since machine learning is used to detect rare species and make conservation and management recommendations, it is important to understand how the models are learning. 

The contribution of this work is as follows: 1) this paper presents an approach based on CNN for bioacoustics signal classification and presents an ensemble XAI approach by combining Grad-CAM and DeepLIFT, 2) identification of sub-population variability within each class/variant through latent space analysis and explanation, and
3) demonstration of XAI's potential to uncover biologically meaningful patterns and generate new scientific hypotheses.

% \textcolor{red}{revise this..DONE}
The remainder of this paper is organized as follows. Section~\ref{sec:related_work} covers the prior works related to XAI and bioacoustic signals. Section III describes the data collection process, including the preparation of bird song recordings from various geographical locations, details on the model training procedure, and the proposed ensemble XAI approach.  Section IV presents the experimental results and analysis, highlighting key findings from the classification and explanation stages. Finally, Section V concludes the paper, summarizing the contributions and suggesting directions for future research.

\section{Background Study}{\label{sec:related_work}}
% \textcolor{red}{1. Remove those refs that are deeply related to bio-acoustics. DONE}

% \textcolor{red}{2. How does these refs related to what you have done?, otherwise this background study will not strengthen the paper. DONE}
% Passive acoustic monitoring (PAM) has been used within the field of ecology for conservation applications, including density estimation and biodiversity monitoring \cite{Perez2021}. Given the sheer amount of data output from automated recording units, ecologists have increasingly been employing machine learning techniques to analyze this data. There are a multitude of applications of machine learning to bioacoustics data to detect the presence or absence of species from their vocalizations within audio data. This type of analysis can paint a picture of an ecological soundscape, which contributes to an understanding of ecosystem health \cite{NietoMora2023}.

%Bioacoustics, the study of sounds made by living things, has witnessed a transformative shift with the advent of artificial intelligence (AI) and machine learning (ML). Traditional methods of manual annotation and spectral analysis are often time-consuming and subjective, limiting the scale and scope of research.
Artificial intelligence (AI) and machine learning (ML) approaches have been used for automated detection, classification, estimating the size of an animal’s vocal repertoire, and even the interpretation of bird vocalizations \cite{piczak2015environmental, stowell2016automatic, kahl, Keen2021}. The application of ML models, such as CNNs and recurrent neural networks (RNNs), has demonstrated remarkable success in these tasks \cite{stowell2016automatic, salamon2017deep, kahl, Liu2022}. Methods like Support Vector Machines (SVMs) and Random Forests have also been utilized \cite{Fagerlund2007}. 
Deep CNNs are sophisticated models that are well-suited to the complex task of classifying bird song. However, they are also \textit{black boxes} and can make biased decisions without users knowing~\cite{Stowell2022}, especially since many biologists do not have computer science domain expertise. 

Because ML is being used to inform conservation efforts, affecting policy-making, it is important that these models become more understandable to the scientists interpreting their predictions. 
% The \textit{black box} nature of many deep learning models poses challenges in understanding the underlying mechanisms and output transparency. 
XAI techniques are crucial for addressing this issue \cite{ arrieta2020explainable}, seeking to achieve interpretability and explainability of complex models' decisions. By providing insights into the decision-making process of ML models, XAI techniques can enhance transparency and trust in automated bioacoustic analyses. XAI techniques like saliency maps and feature importance analysis can reveal which acoustic features are most critical for species identification, offering valuable biological insights \cite{selvaraju2017gradcam, ribeiro2016Lime, sundararajan2017axiomatic}.
% The integration of XAI with AI and ML in bioacoustics enables researchers to validate model predictions against established biological knowledge. The application of XAI in bird song analysis is not merely about improving model accuracy, but also about fostering a deeper understanding of avian behavior and ecology. Moreover, the integration of multimodal data, such as visual and environmental information, can further enhance the accuracy and interpretability of bird song analysis \cite{Rajput}. 

There is an inherent trade-off between model interpretability and performance in complex tasks, and many methods have been developed to obtain post-hoc explanations of a model’s predictions \cite{arrieta2020explainable}. Prominent among these XAI techniques are the model-agnostic approaches LIME and SHAP, and the CNN model-specific approaches Grad-CAM and DeepLIFT. LIME \cite{ribeiro2016Lime} approximates black-box models locally using an interpretable surrogate, such as a linear regression model, by perturbing input data and analyzing the predictions. It offers flexible, localized explanations. SHAP \cite{lundberg2017shap} applies cooperative game theory to compute feature importance by averaging their contributions across all possible combinations. It ensures consistent and globally valid attributions. 

Different from these techniques, DeepLIFT \cite{shrikumar2017deeplift} propagates relevance scores backward through the network, attributing model decisions to input features and aiding in understanding neural network decision-making. Grad-CAM \cite{selvaraju2017gradcam} generates visual heatmaps by computing the gradients of the target class with respect to feature maps, highlighting important regions in input images for CNN predictions. These XAI techniques can be used to debug the model and detect bias in the dataset. They could also possibly be used to generate testable hypotheses~\cite{Lipton2017}. 

When the insight provided by post-hoc explanations is combined with the background knowledge of domain experts, new understanding can emerge~\cite{Jeyasothy}. A review of AI in landscape ecology listed XAI as one of six main opportunities for expansion of landscape ecology methods~\cite{Frazier2024}. LIME has already been used within landscape ecology to make species distribution models more interpretable~\cite{Ryo2021}. Within bioacoustics, SHAP has been used alongside random forest models and clustering analyses to assess soundscapes of shallow marine environments~\cite{Parcerisas2023}. Das~\cite{Das2024} used LIME and SHAP to improve a model for classifying bird species by song and showed that SHAP performed slightly better. 
% Stowell~\cite{Stowell2022} reviewed the use of deep learning in bioacoustics, emphasizing species identification, environmental monitoring, and automated call classification. The paper highlighted the challenge of data scarcity and the need for interpretability in deep learning models. 
Schäfer-Zimmermann~\cite{schäfer2024animal2vec} introduced a self‑supervised transformer for rare‑event detection in bioacoustic data and presented a large‑scale benchmark dataset, noting that their model provides interpretable outputs relevant for ecological research. Govindaprabhu~\cite{govindaprabhu2025bridging} explored the integration of XAI with Convolutional Interactive Learning Neural Networks (CILNN) to predict wildlife behavior through acoustic signals. SHAP was used with CNN to detect audio signals of dangerous animals and prevent human-animal conflict. The authors emphasize the importance of transparency in AI models to enhance ecological applications and conservation efforts. AudioProtoNet~\cite{audioprotopnet2024} introduces an interpretable deep learning approach to bird sound classification by leveraging prototype learning, making model decisions more transparent and explainable.

Inspired by the above endeavours and open challenges, we conducted our research for a comparative analysis between LIME, SHAP, Grad-CAM, and DeepLIFT in explaining bird song classification.

\begin{figure*}[t]
    \centering
    \includegraphics[width=18cm, height=3.5cm]{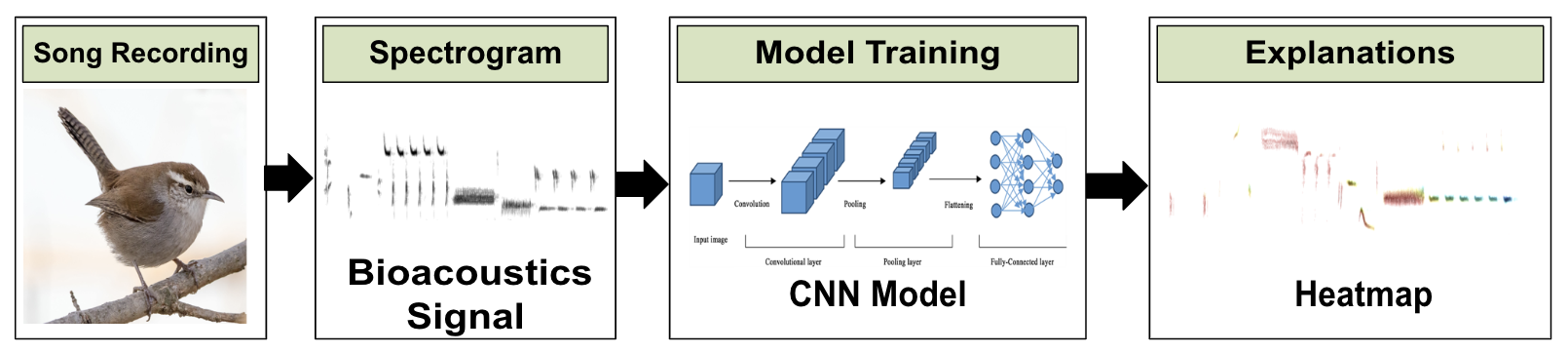}
    \caption{An overview of the pipeline including song recording, classification, and explanation. Songs are transformed into spectrograms and used to train a CNN model. Then, visual explanations, highlighting the regions that most influenced the model’s predictions, were generated.}
    \label{fig:overview}
\end{figure*} 

\section{Methodology}

This section presents data collection, model training, and the ensemble XAI approach.
Figure~\ref{fig:overview} shows an overview of the approach. The process begins with recording bird sounds, which are converted into spectrogram images. A CNN model is trained on these spectrograms for classification, and different XAI techniques: LIME, SHAP, Grad-CAM, and DeepLIFT are employed to generate visual explanations highlighting important regions influencing the model's predictions. 

% To analyze some apparent anomalies in the explanation images, we employed PCA and t-SNE methods to reduce the features of each sample's latent CNN representation and plotted those graphically to visualize their distribution and correlation.

% \textcolor{red}{more details needed here.}

\subsection{Data Collection}
We recorded singing Bewick's wrens, a species of song-learning bird~\cite{kroodsma1974song}, in Arizona and New Mexico in May and June 2023 using a MixPre 3 recorder and a Sennheiser ME62 microphone with a parabolic dish. The sampling frequency was $48000$ Hz, and the bit rate was 24. Each full recording was segmented into 4-second clips, with each clip containing the entirety of a single song. The resulting audio files were bandpass filtered from $1.5$ to $9$ kHz before conversion into spectrogram images~\cite{seewave2008} using a Short-Time Fourier Transform (STFT) with a window length of $512$ samples, $95\%$ overlap, and a dynamic range from $-40$ to $5$ dB. The STF transformation divides the signal into short, overlapping time windows, allowing both time and frequency information to be captured simultaneously. A spectrogram visualizes an audio signal as a two-dimensional image, with time on the x-axis, frequency on the y-axis, and color as the amplitude/intensity of the sound. To preserve fine-grained details in the images, spectrogram image resolution was maintained at \( 480 \times 960 \) pixels.

% Figure~\ref{fig:example_spectrogram_image} shows an example of a spectrogram image from a song recording of a Bewick's Wren singing Eastern variant songs. The spectrogram displays distinct shapes (often referred to as notes, elements, or traces) that correspond to individual sound events within the song. Researchers can use these visual patterns to interpret acoustic structure and classify bird songs. ~\ref{fig:example_spectrogram_image}.   

\subsection{Model Training and Explanation}
In total, 1854 training and 806 test samples were generated from two song variants of \textit{Thryomanes bewickii} (Bewick’s Wren) and a CNN model was trained in a supervised learning manner. CNN architecture was optimized through hyperparameter tuning to achieve high classification accuracy. The final architecture comprised five convolutional layers with kernel size \( \mathcal{K}_1 = 3 \times 3 \), padding \( \mathcal{P} = 1 \), and stride \( \mathcal{S}_1 = 1 \), with output channels of $16$, $32$, $64$, $128$, and $256$ respectively. Each convolutional layer was followed by a ReLU activation and a max-pooling layer with size \( \mathcal{K}_2 = 2 \times 2 \) and stride \( \mathcal{S}_2 = 2 \). The extracted features were flattened and passed through three fully connected layers containing $512$, $1024$, and $2$ neurons. The model was trained using the Adam optimizer with a learning rate \( \eta = 0.001 \) and weight decay \( \lambda = 10^{-5} \). Cross-entropy loss was used as the objective function, and training was performed for 20 epochs with a batch size of 64. 
% The train and test plots are seen in figure ~\ref{fig:train_loss}.

\begin{figure}[t]
    \centering
    \fbox{\includegraphics[width=0.9\linewidth, height=3.7cm]{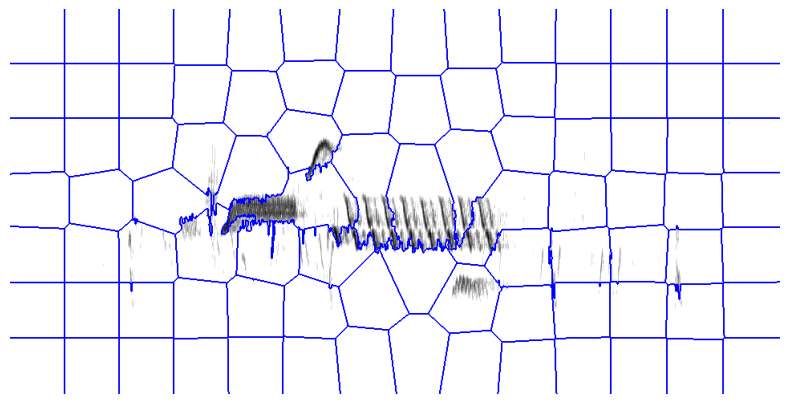}}
    \caption{A test sample divided into 100 super-pixels. 1000 perturbed samples were generated by turning on and off these super-pixels randomly, and the importance of each super-pixel is computed. Note that most of the super-pixels are empty and do not contain any part of the signal.}
    \label{fig:slic}
\end{figure}

% \subsection{XAI Techniques}
% \textcolor{red}{more details needed here. DONE}
For interpretation technique, we began with LIME, a widely recognized and foundational model-agnostic technique that introduced the use of local surrogate models to explain black-box predictions.
% \textcolor {red}{(...why you are saying LIME is simple?} 
LIME divide the image into different segments, known as superpixels, using the SLIC (Simple Linear Iterative Clustering) algorithm. In this work, the image was divided into $100$ segments/super-pixels (as shown in Figure \ref{fig:slic}) to strike a balance between granularity and interoperability. Finer segmentation allowed more detailed explanations, while coarser segmentation simplifies the visual interpretation. To generate local explanations, LIME created $1000$ perturbed samples by randomly turning superpixels on or off and observing how these changes influenced the model’s predictions. It then highlighted the superpixels that contributed most positively to the predicted class, focusing on the top few influential regions. However, LIME is a local explanation technique and aims to explain the model's behavior in the local neighborhood of the given test sample and thus is not generalizable.

To get a stable and global explanation, we also investigated SHAP, however, due to its high computational cost, especially for high-resolution images like ours ($480\times960$ pixels), only $50$ background samples could be used to approximate SHAP values effectively. These background samples served as a reference for interpreting how each feature contributes to the model's predictions. The resulting global attribution mask was then consistently applied to each image to highlight important regions.

As the last two methods were model-agnostic, though the model was available, we then wanted to leverage the knowledge of model architecture. We implemented Grad-CAM~\cite{selvaraju2017gradcam}, which retrieves the gradients and feature maps from the model’s last convolutional layer. The mean gradients were pooled across channels, and each feature map was weighted by its corresponding gradient importance. A class-discriminative heatmap was generated by computing a weighted average of the feature maps, followed by the application of a Rectified Linear Unit (ReLU) to retain only positive contributions. The resulting heatmap was normalized by applying min-max scaling, ensuring all values lie within the $[0, 1]$ range:
\[
H_{\text{norm}} = \frac{H - \min(H)}{\max(H) - \min(H) + \varepsilon}
\]
where $H$ is the original heatmap and $\varepsilon$ is a small constant added for numerical stability. 
% For visualization, this heatmap was overlaid on the original grayscale spectrogram, allowing class-relevant regions to be highlighted without obscuring background signal structures. The computed heatmap was resized to match the input spectrogram’s dimensions for accurate spatial alignment before overlay. The heatmap was rendered using the jet colormap, which assigns cooler colors (blue) to low-importance regions, transitions through green and yellow, and emphasizes high-importance areas in red. This color gradient effectively conveys the spatial distribution of model attention and highlights the regions that most strongly contributed to the model’s prediction.
Lastly, we employed the DeepLIFT \cite{shrikumar2017deeplift}  technique, which attributed the model’s prediction back to the input features. This method computes contribution scores by comparing the activations of each neuron to a reference input, enabling the identification of features that significantly influenced the model’s decision. In our implementation, a plain white image was used as the reference baseline, and we applied a ReLU operation to the resulting channel-wise relevance scores to retain only positive contributions. For visualization, the relevance map was normalized, resized to match the original spectrogram dimensions, and overlaid on the grayscale input using the same jet colormap as Grad-CAM. This consistent color mapping facilitates direct visual comparison across the explanation techniques. 
% \textcolor{red}{How Grad-CAM and DeepLIFT are different and what different information they can provide. Do they complement each other or provide same information. LIKE LIME and SHAP are different and often used for different types of explanation....}

% \begin{figure}[t]
%     \centering
%     \includegraphics[width=1\linewidth, height=4cm]{images/loss-accuracy.png}
%     \caption{Training loss and Testing accuracy for white, colored(black), and mixed datasets}
%     \label{fig:train_loss}
% \end{figure}

\begin{table}[t]
\centering
\caption{Classification Performance across Different Backgrounds}
\begin{tabular}{l|c|c|c|c}
\hline
Background & Accuracy & Recall & Precision & F1 Score \\
\hline
Black & 93.18\%  & 0.9438 & 0.9628 & 0.9532 \\
White & 94.00\%  & 0.9522 & 0.9658 & 0.9590 \\
Mixed & 94.83\%  & 0.9452 & 0.9839 & 0.9642 \\
\hline
\end{tabular}
\label{tab:model_performance}
\end{table}

\begin{figure*}[t]
    \centering
    % \subfloat[original spectrogram for Eastern class]{\includegraphics[width=5cm, height=3cm]{images/original 0.png}}
    \subfloat[LIME, Eastern class]{\fbox{\includegraphics[width=6cm, height=2cm]{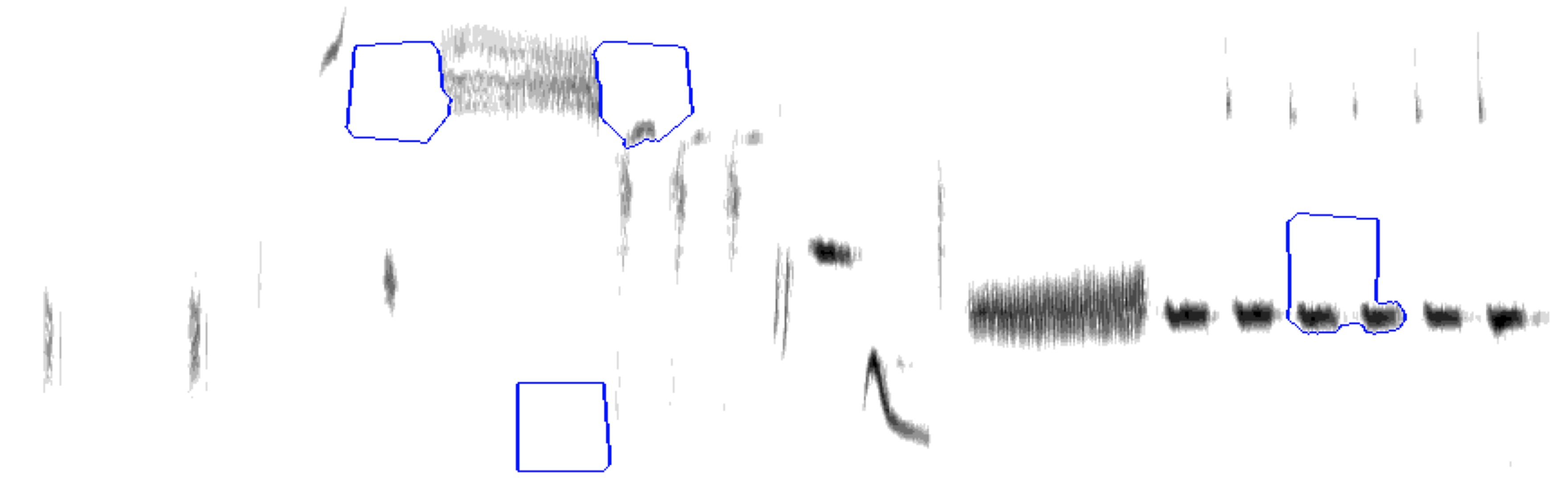}}}\hspace{0.0cm}
    \subfloat[SHAP, Eastern class]{\fbox{\includegraphics[width=6cm, height=2cm]{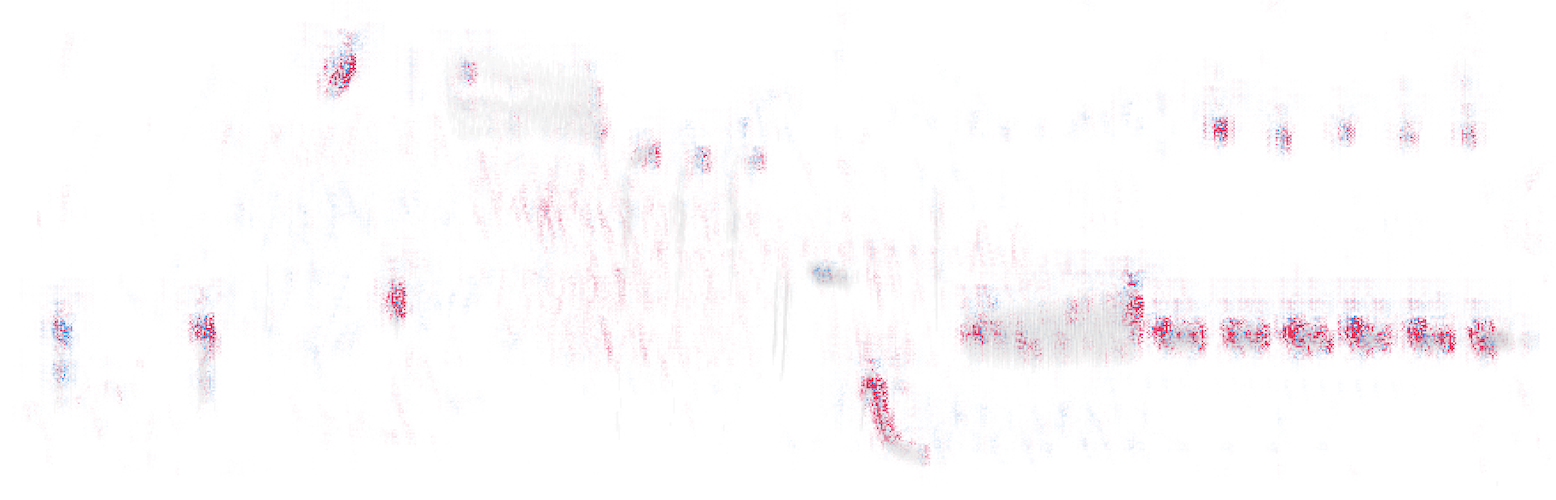}}}\\
    \subfloat[LIME, Mexican class]{\fbox{\includegraphics[width=6cm, height=2cm]{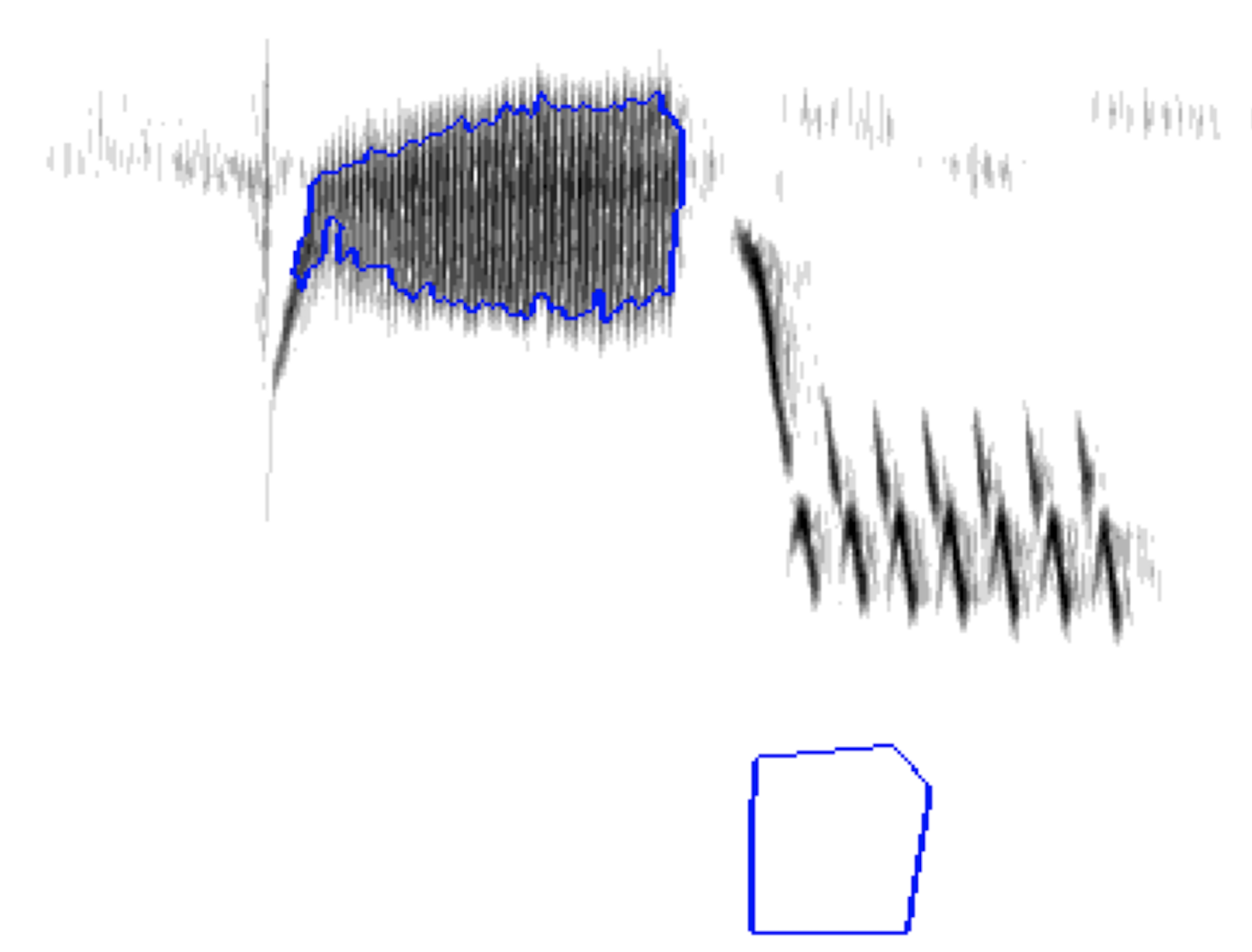}}}\hspace{0.0cm}
    \subfloat[SHAP, Mexican class]{\fbox{\includegraphics[width=6cm, height=2cm]{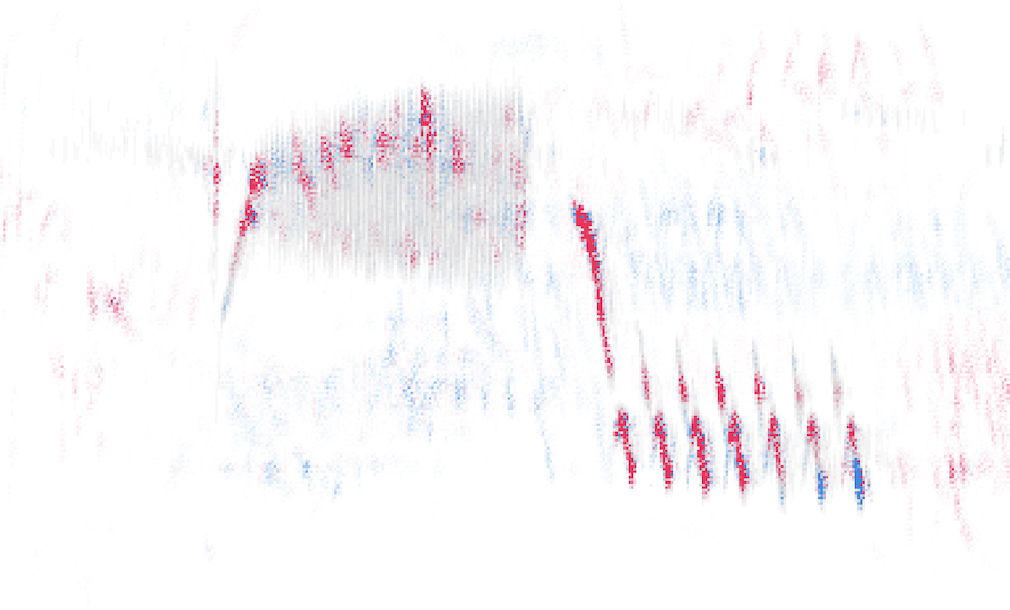}}}
    \caption{ Visual explanations obtained using LIME, and SHAP for two samples, one from the Eastern class and the other from the Mexican class.}
    \label{fig:eastern_mixed_LIME_SHAP}
\end{figure*}

\begin{figure*}[t]
    \centering
    \subfloat[Grad-CAM, Eastern class]{\fbox{\includegraphics[width=4.2cm, height=2cm]{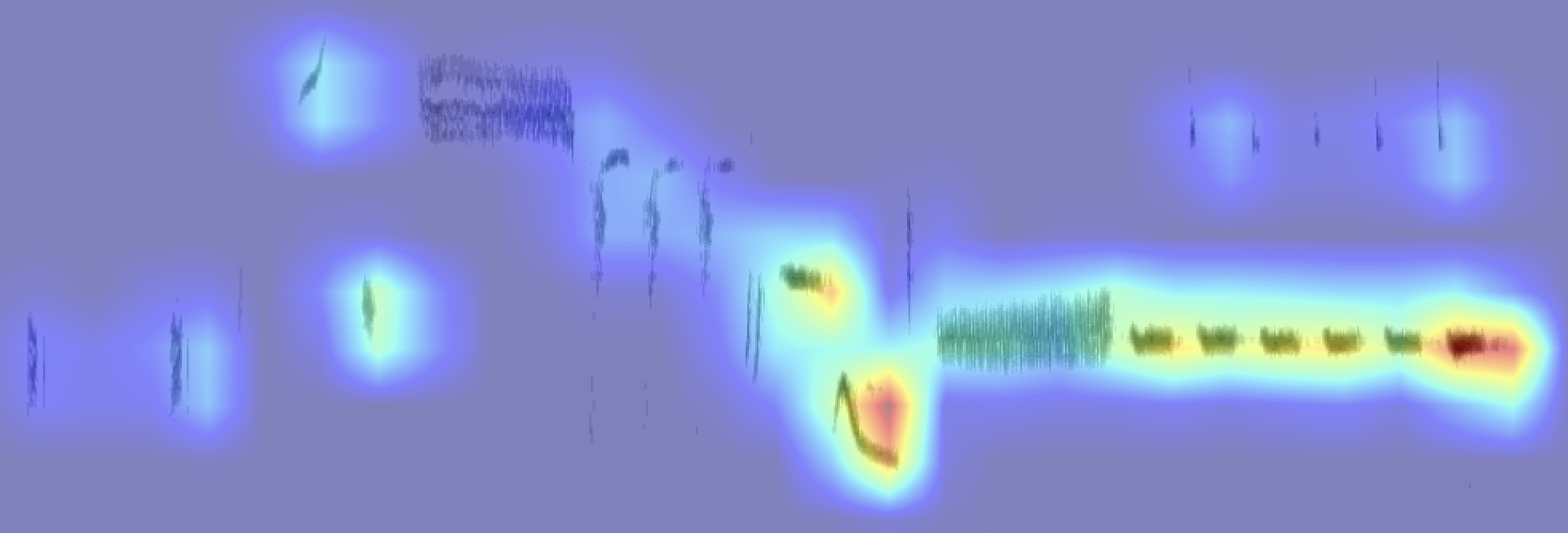}}}\hspace{0.0cm}
    \subfloat[DeepLIFT, Eastern class]{\fbox{\includegraphics[width=4.2cm, height=2cm]{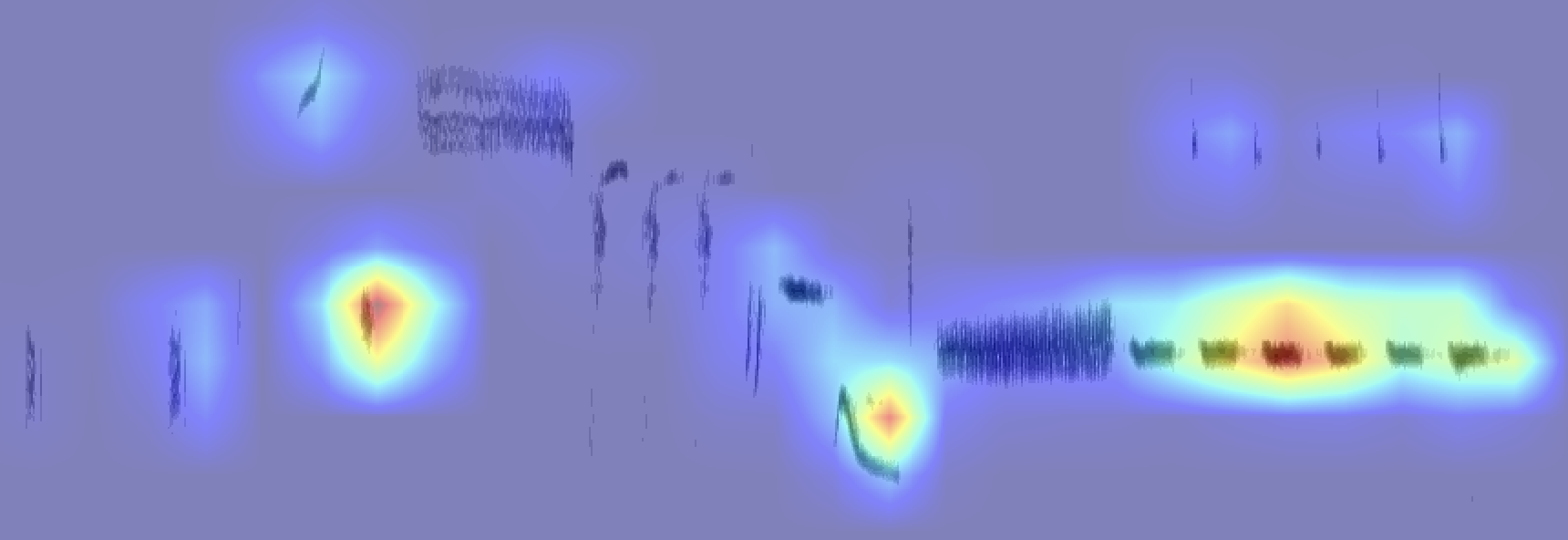}}}
    \subfloat[Weighted Average Heatmap]{\fbox{\includegraphics[width=4.2cm, height=2cm]{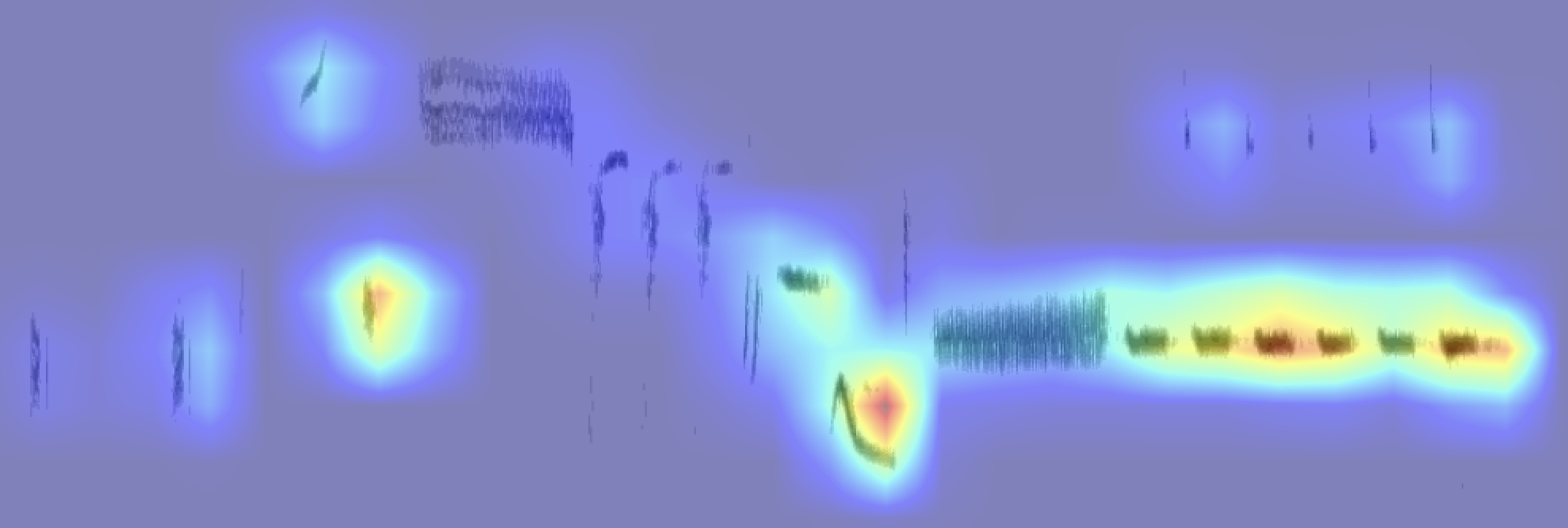}}}
    \subfloat[Element-wise Max Heatmap]{\fbox{\includegraphics[width=4.2cm, height=2cm]{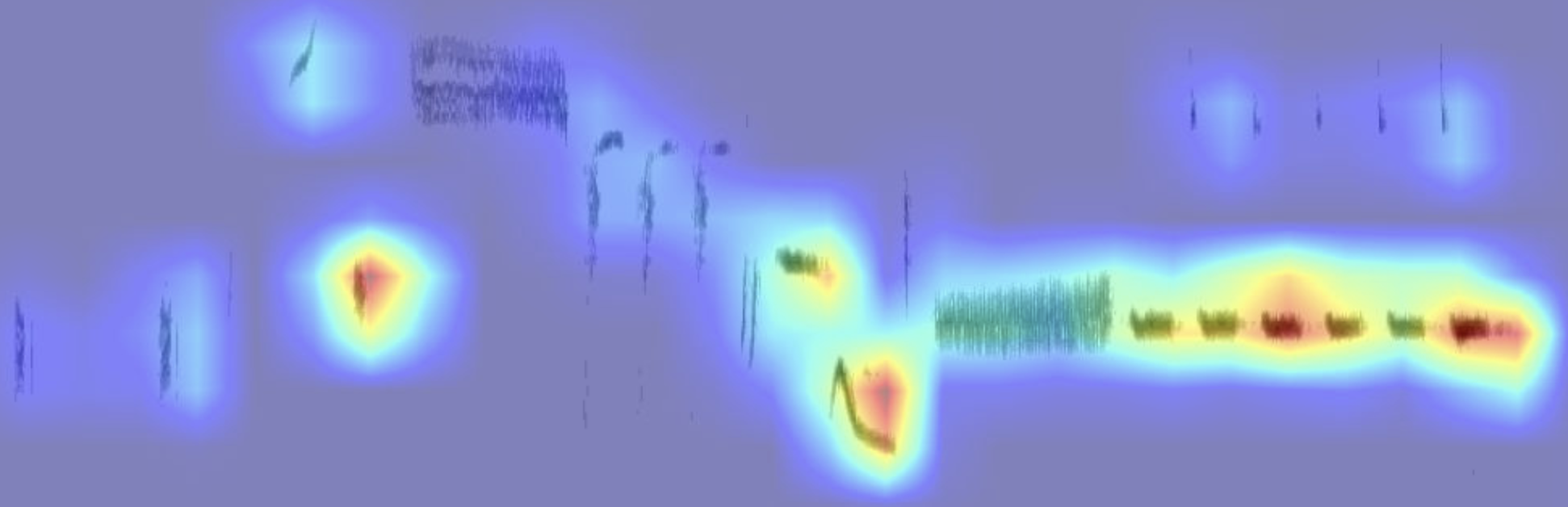}}}
    \hspace{0.0cm}
    \\
    \subfloat[Grad-CAM, Mexican class]{\fbox{\includegraphics[width=4.2cm, height=2cm]{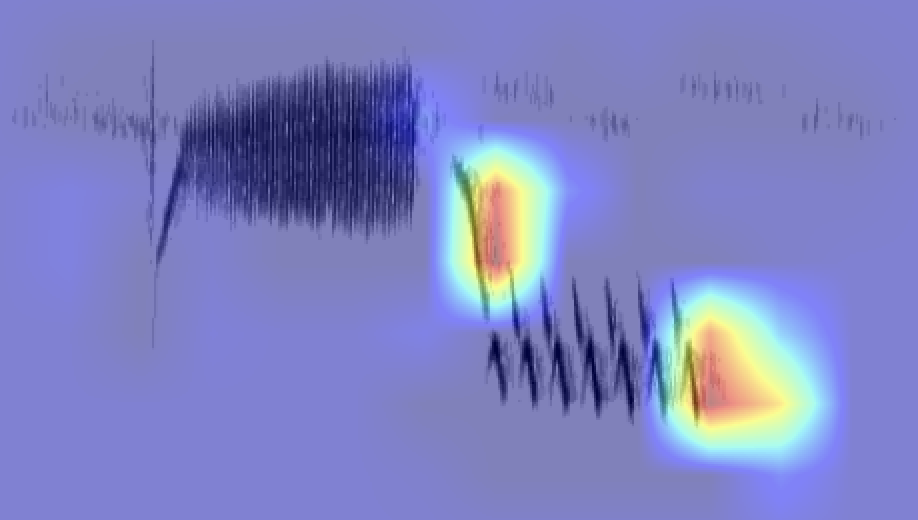}}}\hspace{0.0cm}
    \subfloat[DeepLIFT, Mexican class]{\fbox{\includegraphics[width=4.2cm, height=2cm]{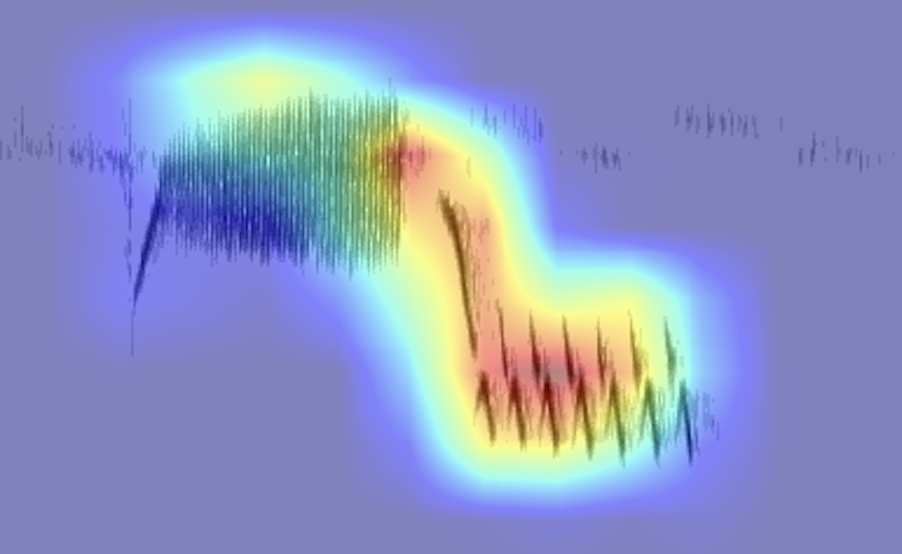}}}
    \subfloat[Weighted Average Heatmap]{\fbox{\includegraphics[width=4.2cm, height=2cm]{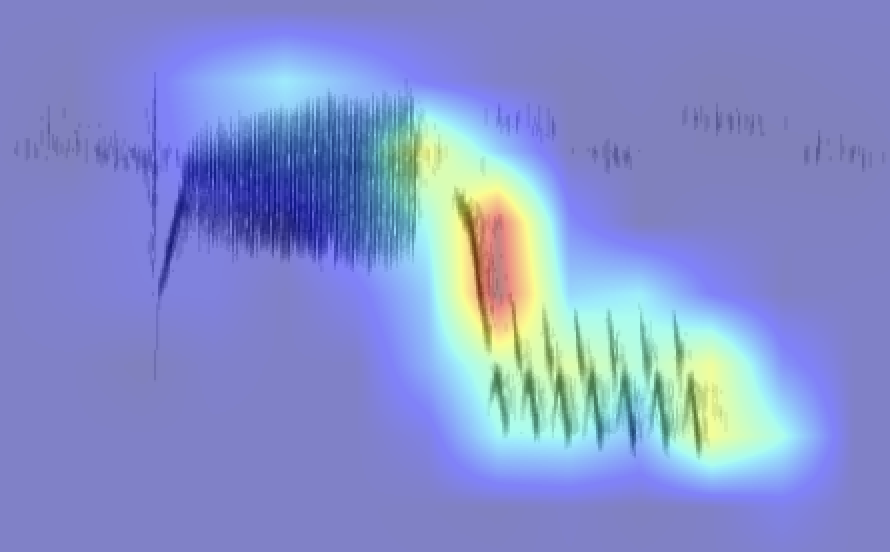}}}
    \subfloat[Element-wise Max Heatmap]{\fbox{\includegraphics[width=4.2cm, height=2cm]{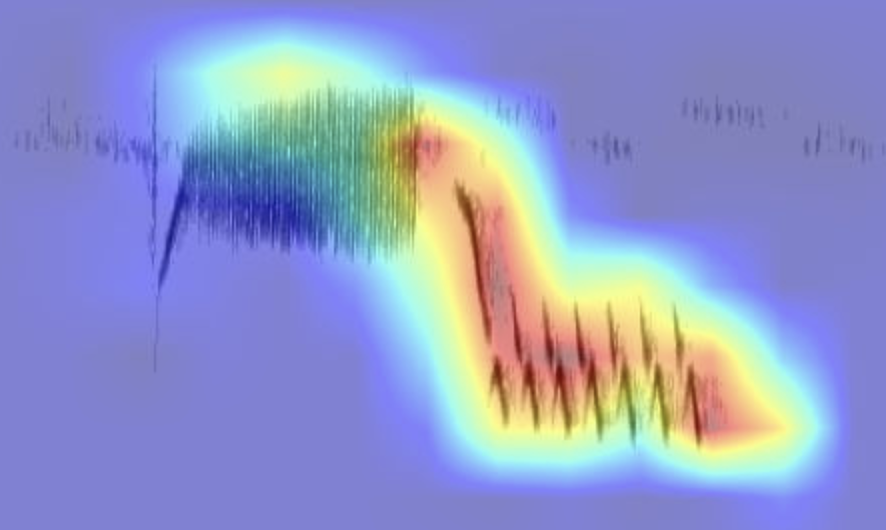}}}

    \hspace{0.0cm}
    \caption{Visual explanations obtained using Grad-CAM, DeepLIFT, and their ensemble heatmaps for the same two samples of Fig \ref{fig:eastern_mixed_LIME_SHAP}.}
    \label{fig:eastern_mixed_Grad}
\end{figure*}

\subsection{Ensemble of Grad-CAM and DeepLIFT Saliency Maps}
While Grad-CAM highlights coarse spatial regions of attention by leveraging gradients in convolutional layers, DeepLIFT provides fine-grained, direction-sensitive attributions by propagating contributions from a reference input.  Although both methods generate heatmaps that highlight class-relevant regions in the input, they rely on fundamentally different attribution principles. 
% Grad-CAM leverages gradient-based activations from convolutional layers, while DeepLIFT utilizes reference-based relevance propagation. 
To enhance the robustness and interpretability of visual explanations, the saliency outputs from Grad-CAM and DeepLIFT were ensembled into unified representations.

To harness the complementary nature of these two approaches, their heatmaps were fused using two simple strategies: (1) \textit{a weighted average} and (2) \textit{element-wise maximum}. Let $H_{\text{CAM}}$ and $H_{\text{LDF}}$ denote the respective saliency maps of Grad-CAM and DeepLIFT, the average-based combined saliency map $H_c$ is computed using equation~\ref{eq:avg_fusion}.
\begin{equation}
H_c = w_1 \cdot H_{\text{CAM}} + w_2 \cdot H_{\text{LDF}}
\label{eq:avg_fusion}
\end{equation}
where $w_1 = w_2 = 0.5$, and element-wise maximum ensemble saliency map is computed using equation~\ref{eq:max_fusion} where $i$ and $j$ represents pixel location.
\begin{equation}
H_{\text{max}}(i,j) = \max\left( H_{\text{CAM}}(i,j), H_{\text{LDF}}(i,j) \right)
\label{eq:max_fusion}
\end{equation}

Both ensemble strategies were applied after individually normalizing each heatmap to the range $[0, 1]$ to ensure consistent scale and dynamic range. These techniques integrate the localized, high-contrast sensitivity of Grad-CAM with the broader, context-aware relevance estimation of DeepLIFT with aim to generate holistic and informative saliency maps.

\section{Results and Discussion}
% \textcolor{red} {Is this repeating previous info?: To interpret the predictions of the trained CNN model, we applied four widely used XAI techniques. LIME and SHAP are model-agnostic methods that generate explanations by approximating the model locally or computing feature attributions based on input perturbations and Shapley values, respectively. In contrast, Grad-CAM and DeepLIFT are model-specific techniques designed to work with CNNs, leveraging internal activations or gradients to highlight salient regions of input contributing to the prediction.} 
By comparing the different XAI techniques, we aim to evaluate not only the quality and consistency of the explanations they provide but also their utility in uncovering biologically meaningful patterns in spectrogram images. This comparative analysis enabled us to identify the most interpretable and reliable explanation strategy for the bioacoustic classification task.

% \subsection{Model Training}
To assess the effect of visual representation on model performance and interpretability, spectrograms were generated with white and black colored backgrounds. Table~\ref{tab:model_performance} presents the classification performance of the CNN model trained on spectrogram datasets with three different background configurations: black, white, and mixed (combining both). The model trained on the mixed-background dataset achieved the highest accuracy (94.83\%), as well as the best precision (0.9839) and F1 score (0.9642), indicating enhanced robustness and generalization. 
% While the white-background dataset slightly outperformed the black one in accuracy and F1 score, the mixed configuration consistently yielded superior results across all metrics. 
These findings suggest that training with visually diverse representations helps the model better capture discriminative features and improves performance.
% on unseen samples.

\subsection{Traditional XAI Explanation Analysis}
% Extensive experiments were conducted, and this subsection presents a comparative analysis. 
After training the model, one test sample was randomly selected from each class or variant, and visual explanations were generated using LIME and SHAP, as shown in Figure~\ref{fig:eastern_mixed_LIME_SHAP} for Eastern class signals (a, b) and Mexican class spectrograms (c, d). LIME's explanations shown in Fig~\ref{fig:eastern_mixed_LIME_SHAP} (a) and (c) highlight blue colored segments which positively contributed to the model’s prediction. 
However, its reliance on random perturbations introduces variability. Repeated runs on the same image produce different highlighted regions. 
As a result, LIME sometimes marked empty or irrelevant regions as important, reducing the reliability of the explanation. In contrast, SHAP explanations, shown in Fig~\ref{fig:eastern_mixed_LIME_SHAP}(b) and (d), outlined some relevant regions (red marked area) of the spectrogram which positively influence the model's predictions. In comparison with LIME, SHAP provides better explanations for the model's decision making; however neither provides strong and conclusive reasoning for the predictions without further analysis.

% \begin{figure*}[t]
%     \centering
%     % \subfloat[original spectrogram for Eastern class]{\includegraphics[width=5cm, height=3cm]{images/original 0.png}}
%     \subfloat[GradCAM, Eastern class]{\fbox{\includegraphics[width=4.2cm, height=2cm]{images/gcam mixed 0.png}}}\hspace{0.0cm}
%     \subfloat[LDF, Eastern class]{\fbox{\includegraphics[width=4.2cm, height=2cm]{images/LDF-0.png}}}\hspace{0.0cm}
%     \subfloat[Combination, Eastern class]{\fbox{\includegraphics[width=4.2cm, height=2cm]{images/combi-0.png}}}\hspace{0.0cm}
 
%     \subfloat[GradCAM, Maxican class]{\fbox{\includegraphics[width=4.2cm, height=2cm]{images/gcam_1_1.png}}}\hspace{0.0cm}
%     \subfloat[LDF, Mexican class]{\fbox{\includegraphics[width=4.2cm, height=2cm]{images/LDF_1_1.png}}}\hspace{0.0cm}
%     \subfloat[Combination, Mexican class]{\fbox{\includegraphics[width=4.2cm, height=2cm]{images/combination_1_1.png}}}\hspace{0.0cm}
%     \caption{ Visual comparison of Grad-CAM, LDF, and their combined heatmaps for two sample spectrograms. The combined saliency maps capture the shared and complementary regions emphasized by both methods, resulting in a more comprehensive visual explanation.}
%     \label{fig:combined_images}
% \end{figure*}

\begin{figure*}[t]
    \centering
    \subfloat[PCA]{\includegraphics[width=7cm, height=5cm]{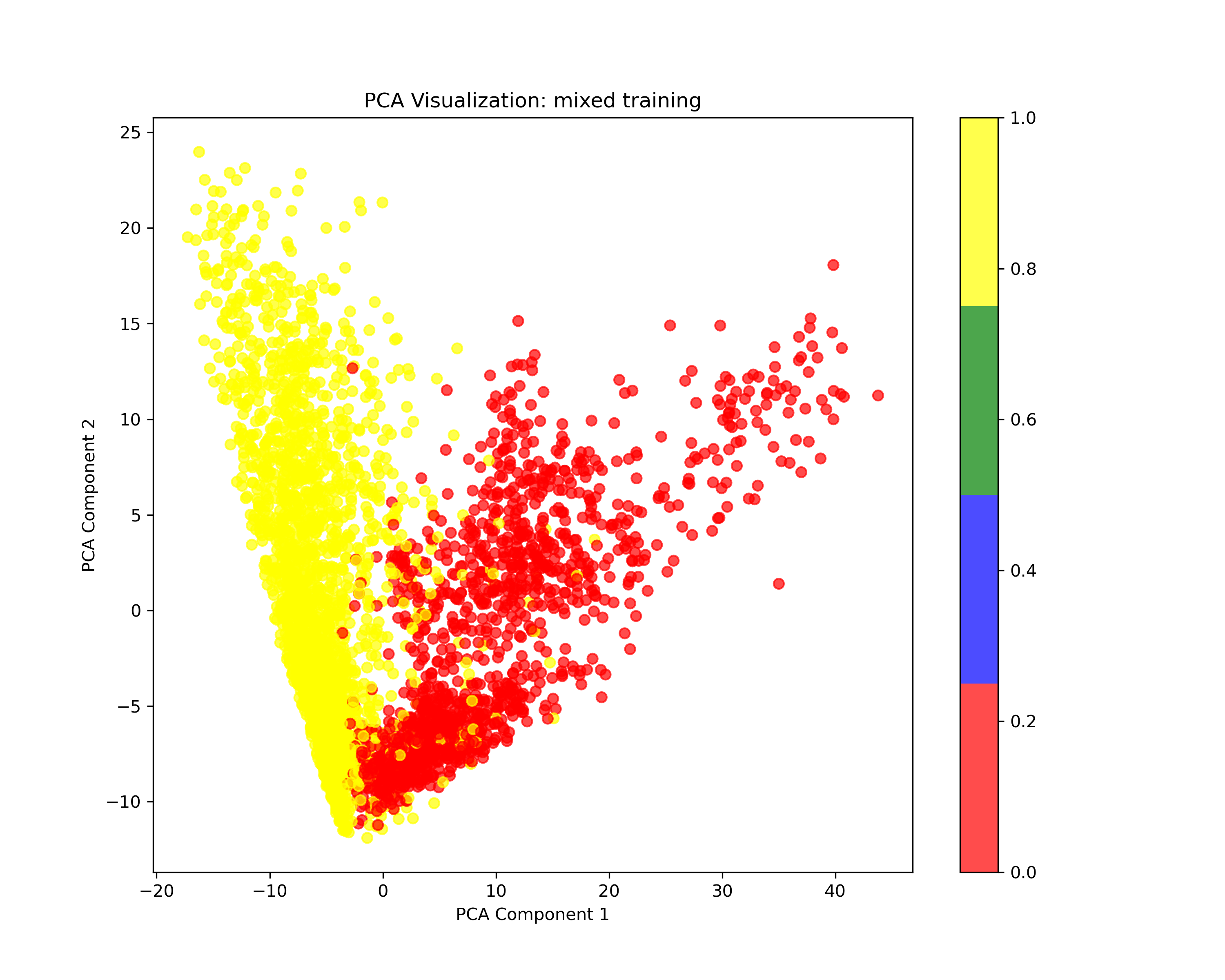}}
    \subfloat[t-SNE]{\includegraphics[width=7cm, height=5cm]{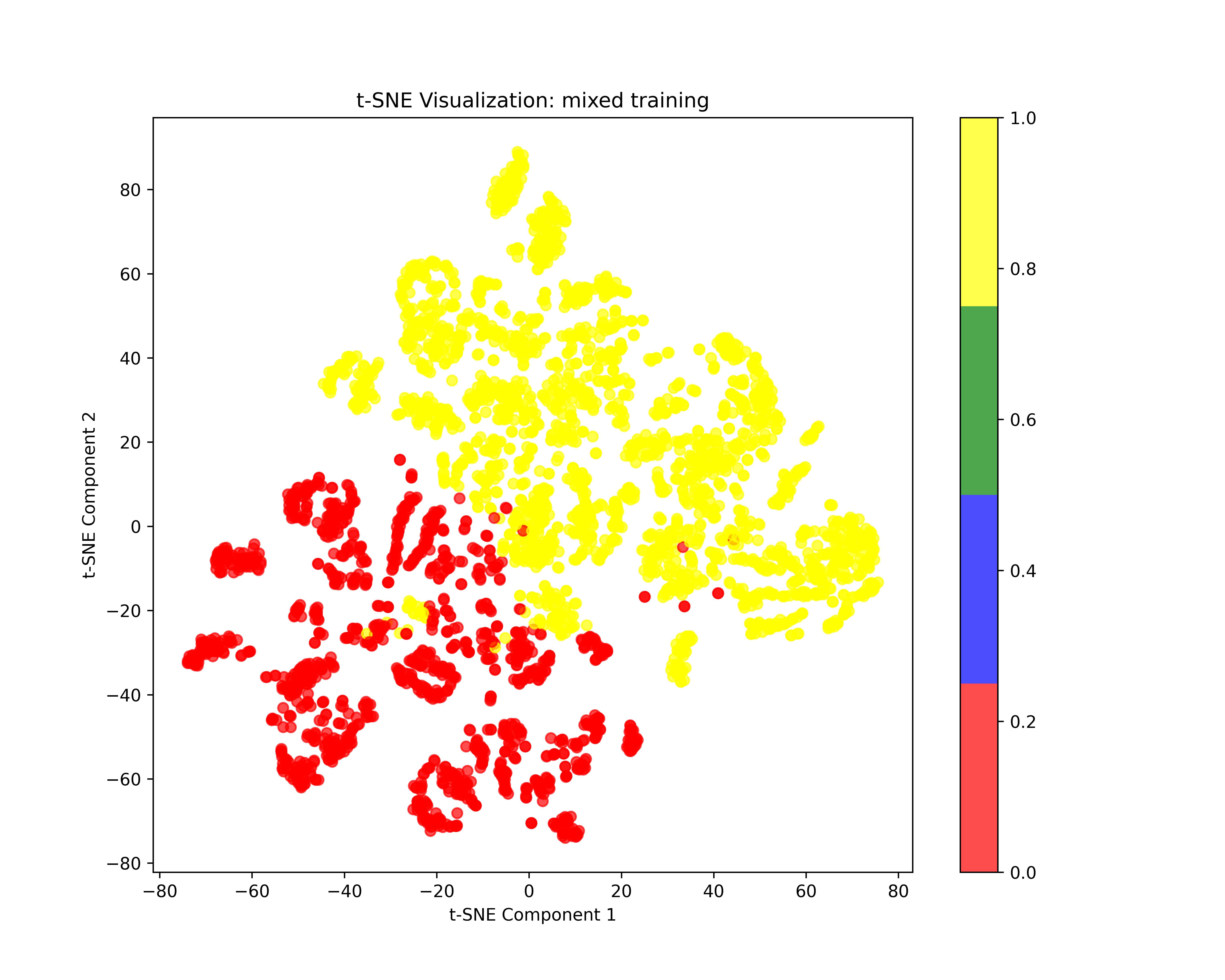}}
    \caption{Distribution of samples in latent space using (a) PCA, and (b) t-SNE, where Eastern and Mexican classes are represented by red and yellow dots. }
    % t-SNE shows several clusters or sub-classes within each class.}
    \label{fig:pca_tsne}
\end{figure*}

\begin{figure}[t]
  \centering
  \includegraphics[width=\linewidth]{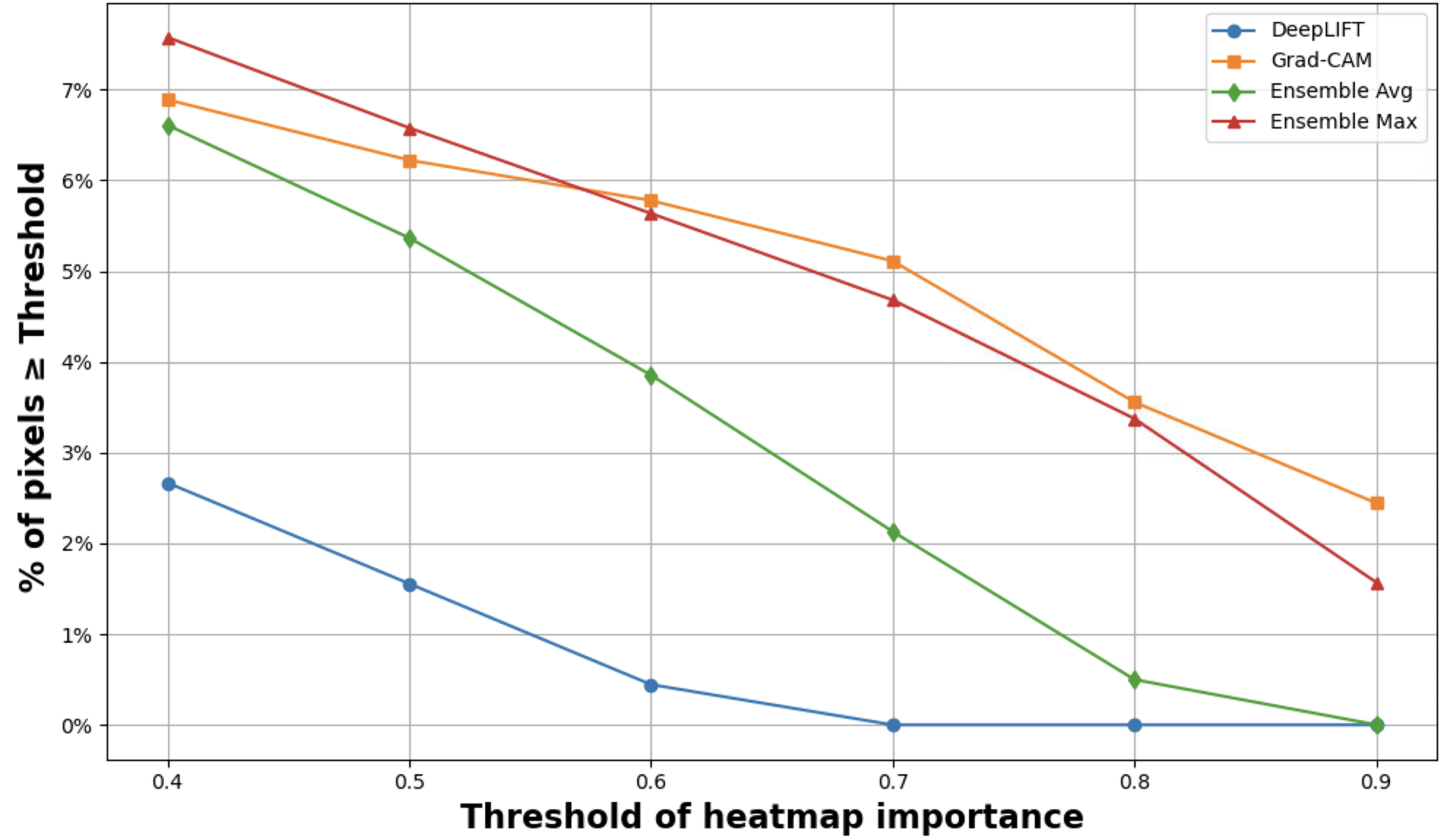}
  \caption{
    Percentage of heatmap pixels exceeding importance thresholds for four saliency maps: DeepLIFT, Grad-CAM, average ensemble, and maximum ensemble. The ``Ensemble Max" consistently highlights more relevant regions across thresholds, indicating it captures a more comprehensive set of discriminative features from both Grad-CAM and DeepLIFT.
  }
  \label{fig:threshold_curve}
\end{figure}

\begin{figure*}[t]
    \centering
    \subfloat[Eastern, E1 sample Grad-CAM]{\fbox{\includegraphics[width=4.2cm, height=2cm]{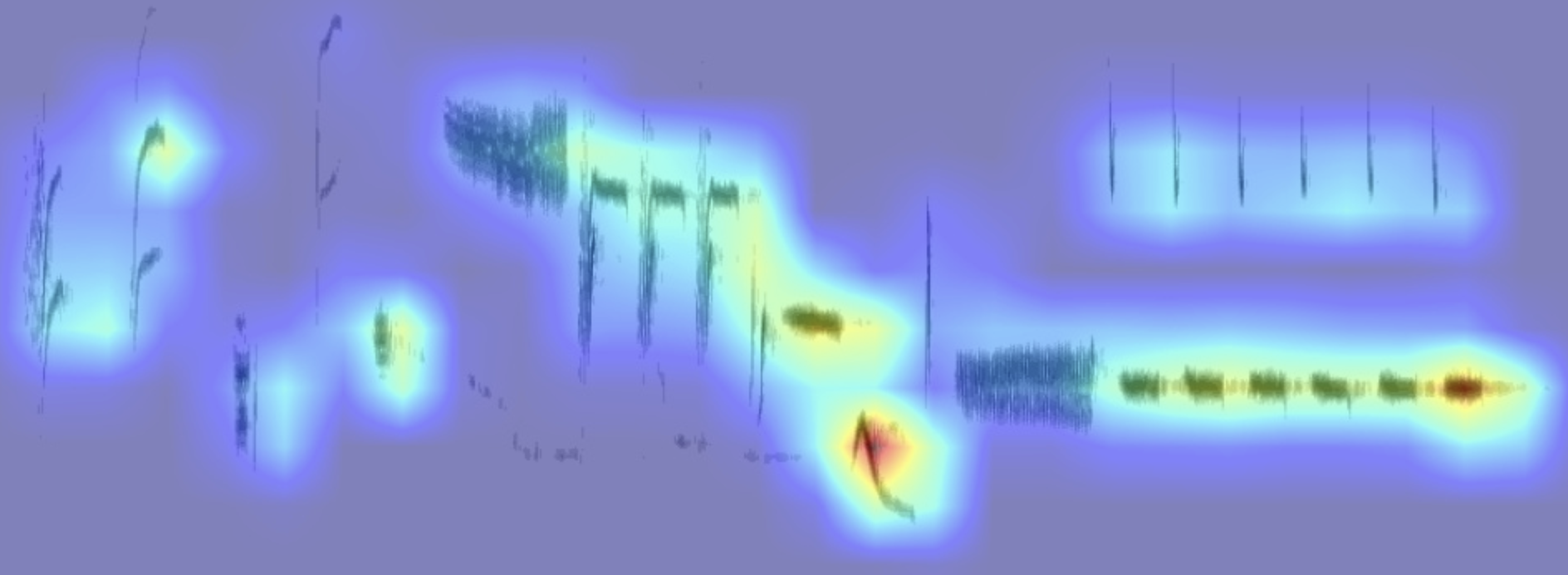}}} \hspace{0.0cm}
    \subfloat[Eastern, E2 sample Grad-CAM]{\fbox{\includegraphics[width=4.2cm, height=2cm]{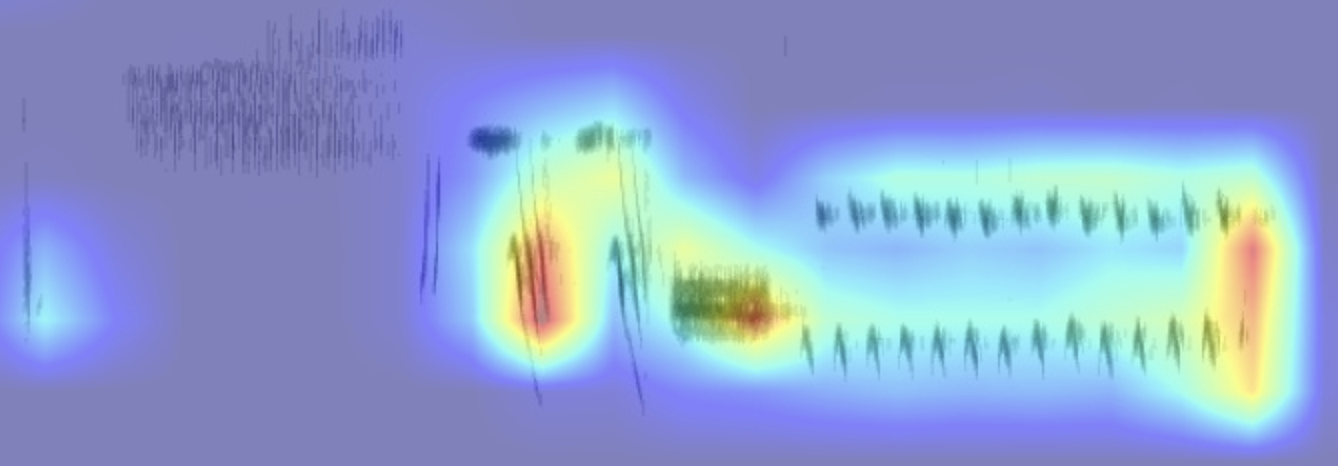}}}\hspace{0.0cm} 
    \subfloat[Eastern, E3 sample Grad-CAM]{\fbox{\includegraphics[width=4.2cm, height=2cm]{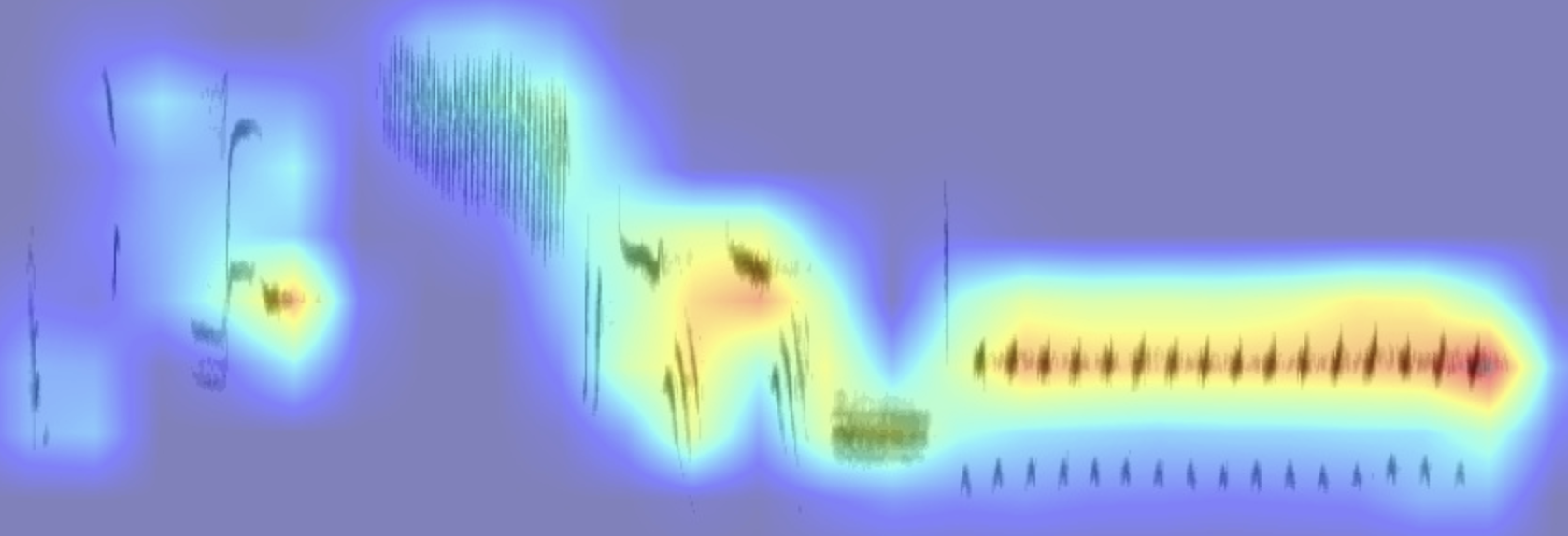}}}\hspace{0.0cm}
    \subfloat[Eastern, E4 sample Grad-CAM]{\fbox{\includegraphics[width=4.2cm, height=2cm]{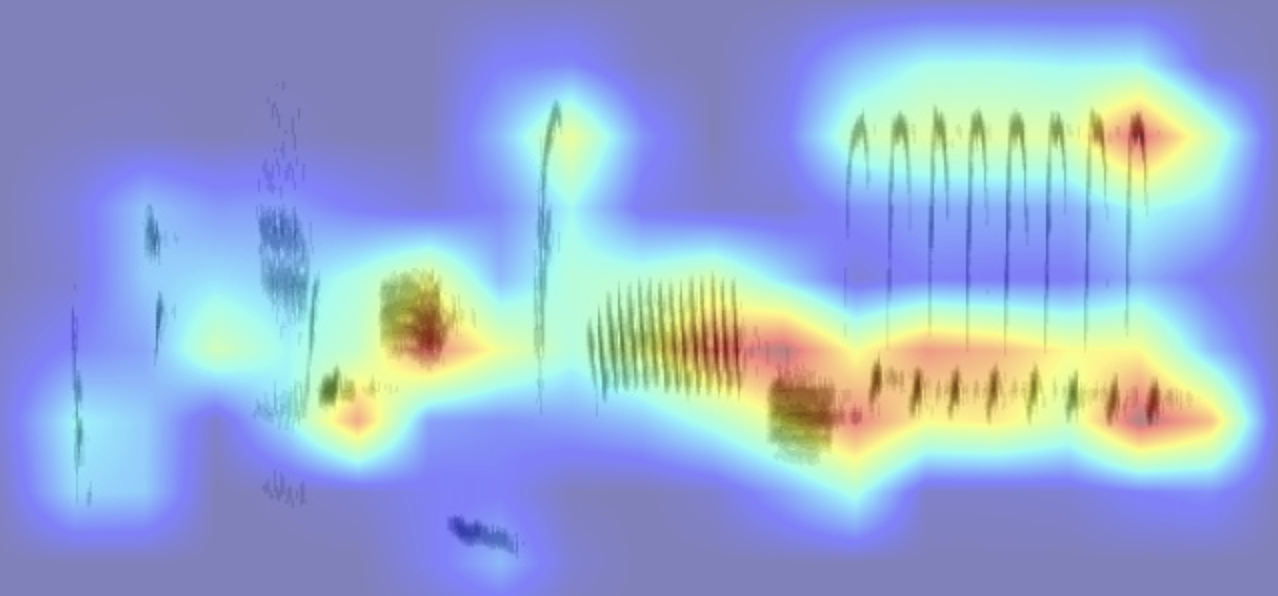}}}\\

    \subfloat[Eastern, E1 sample DeepLIFT]{\fbox{\includegraphics[width=4.2cm, height=2cm]{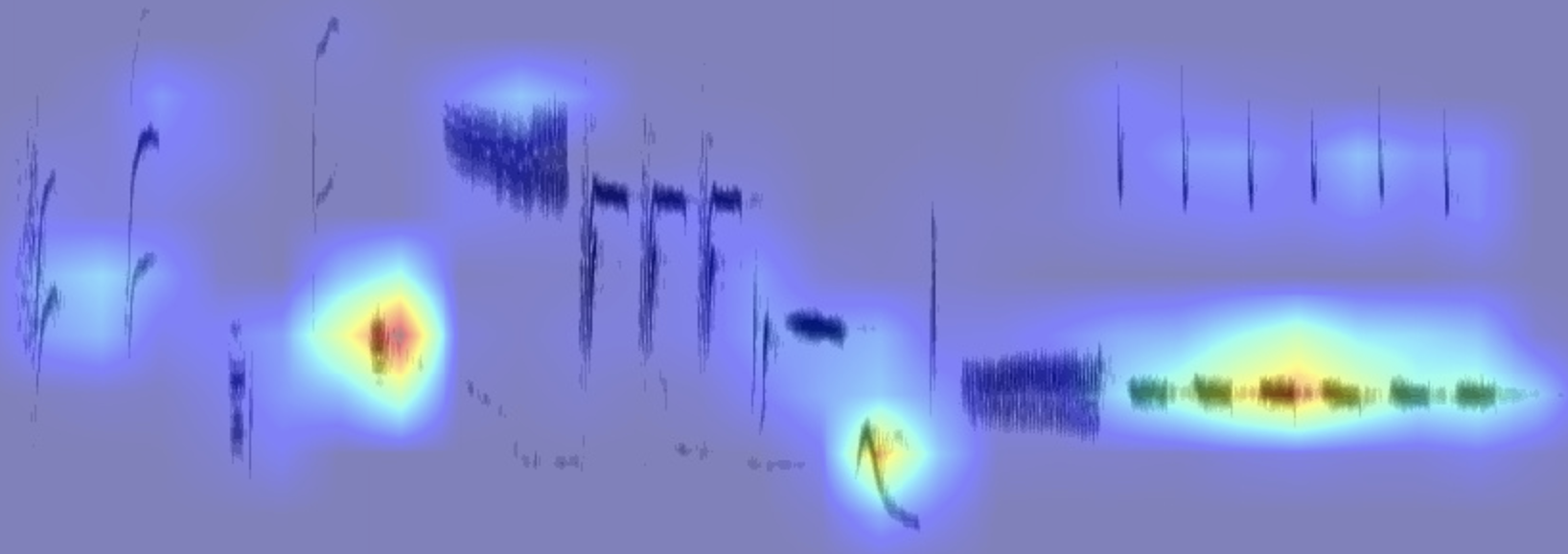}}} \hspace{0.0cm}
    \subfloat[Eastern, E2 sample DeepLIFT]{\fbox{\includegraphics[width=4.2cm, height=2cm]{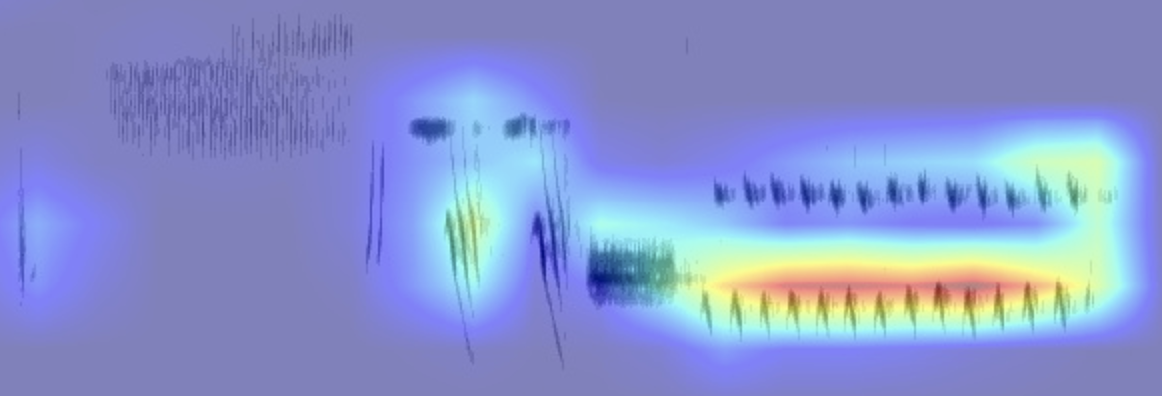}}}\hspace{0.0cm} 
    \subfloat[Eastern, E3 sample DeepLIFT]{\fbox{\includegraphics[width=4.2cm, height=2cm]{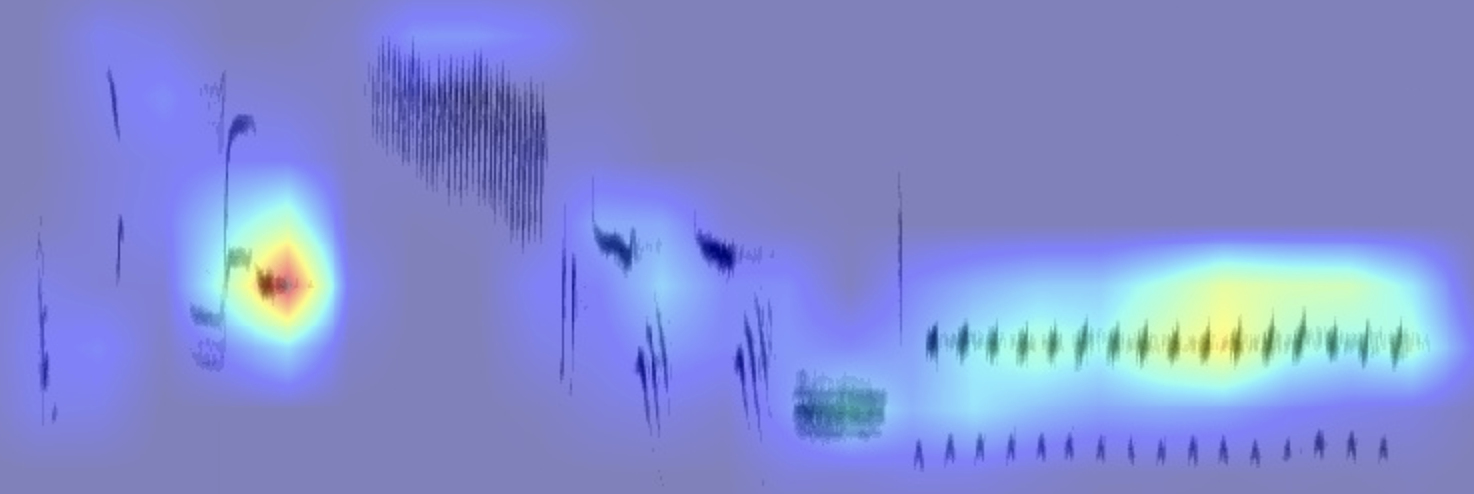}}}\hspace{0.0cm}
    \subfloat[Eastern, E4 sample DeepLIFT]{\fbox{\includegraphics[width=4.2cm, height=2cm]{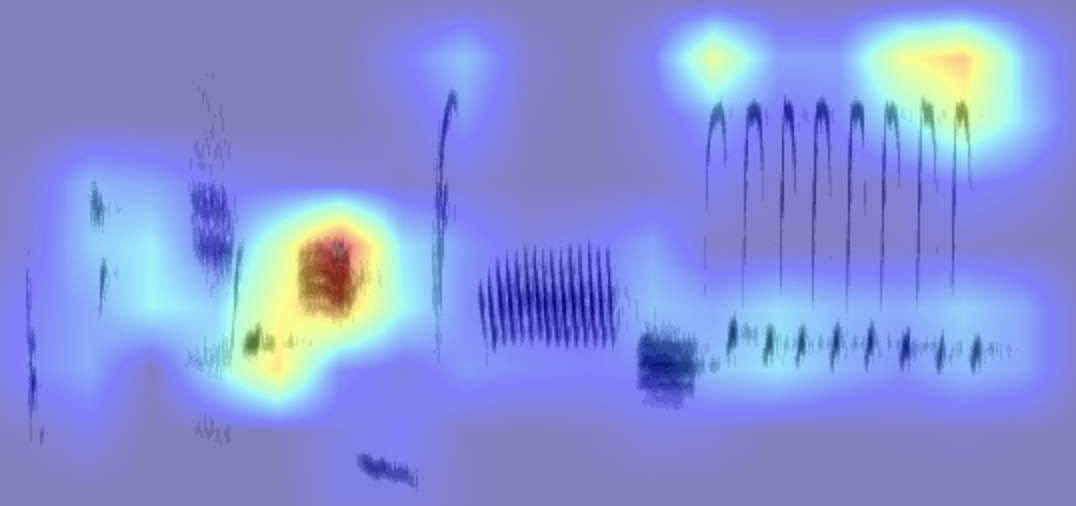}}}\\

    \subfloat[Mexican, M1 sample Grad-CAM]{\fbox{\includegraphics[width=4.2cm, height=2cm]{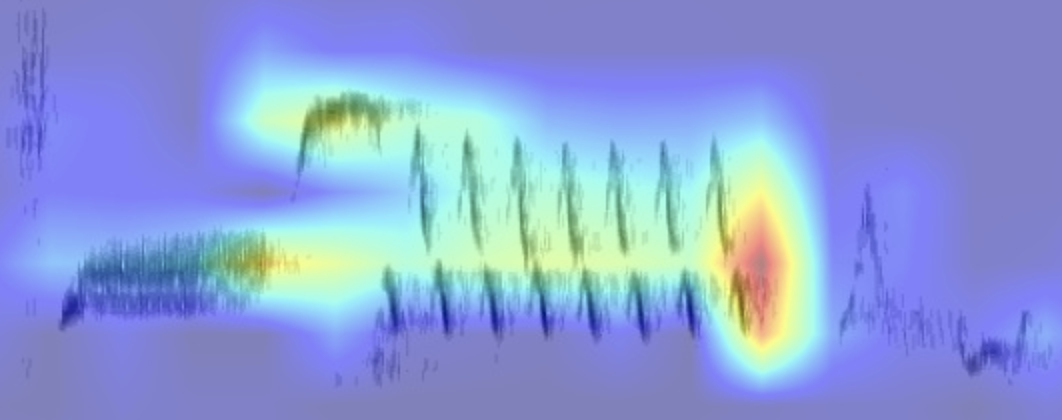}}} \hspace{0.0cm}
    \subfloat[Mexican, M2 sample Grad-CAM]{\fbox{\includegraphics[width=4.2cm, height=2cm]{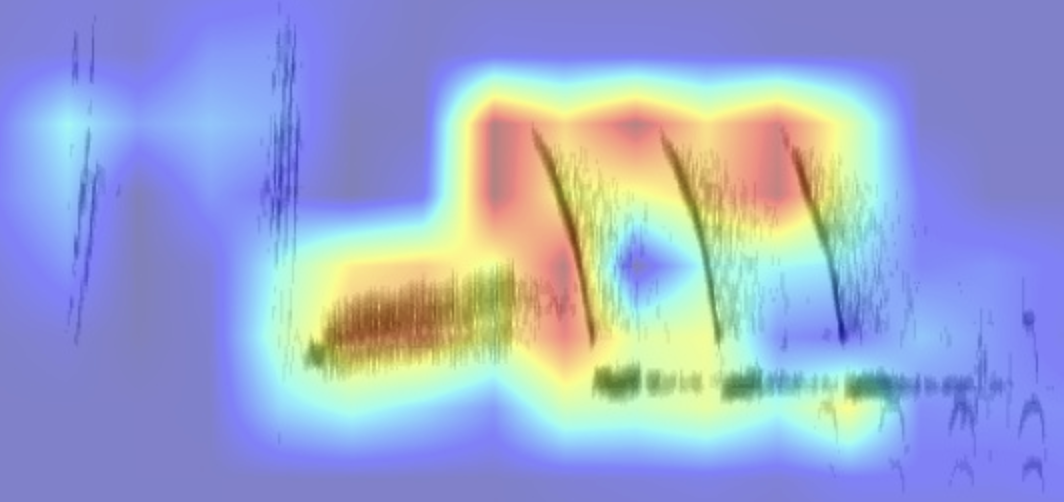}}}\hspace{0.0cm} 
    \subfloat[Mexican, M3 sample Grad-CAM]{\fbox{\includegraphics[width=4.2cm, height=2cm]{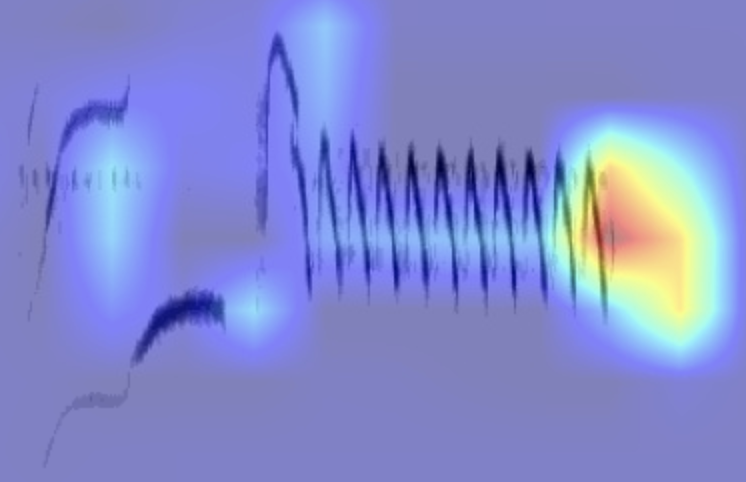}}}\hspace{0.0cm}
    \subfloat[Mexican, M4 sample Grad-CAM]{\fbox{\includegraphics[width=4.2cm, height=2cm]{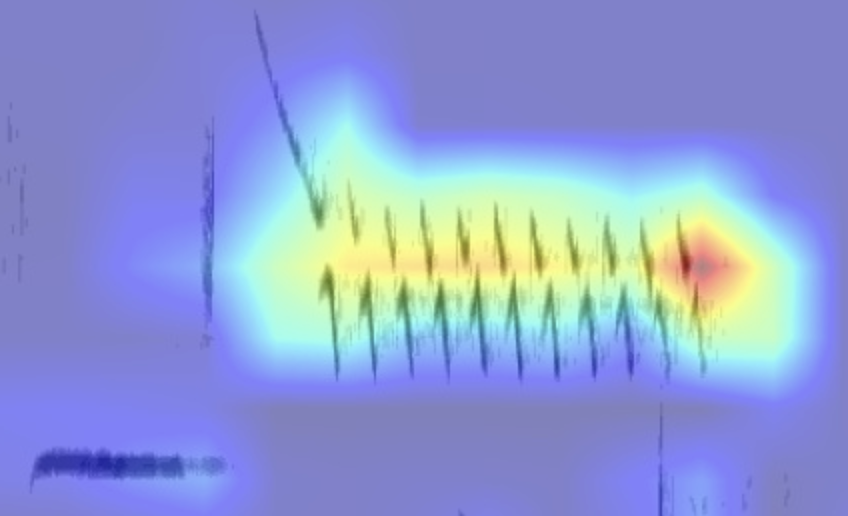}}} \\

    \subfloat[Mexican, M1 sample DeepLIFT]{\fbox{\includegraphics[width=4.2cm, height=2cm]{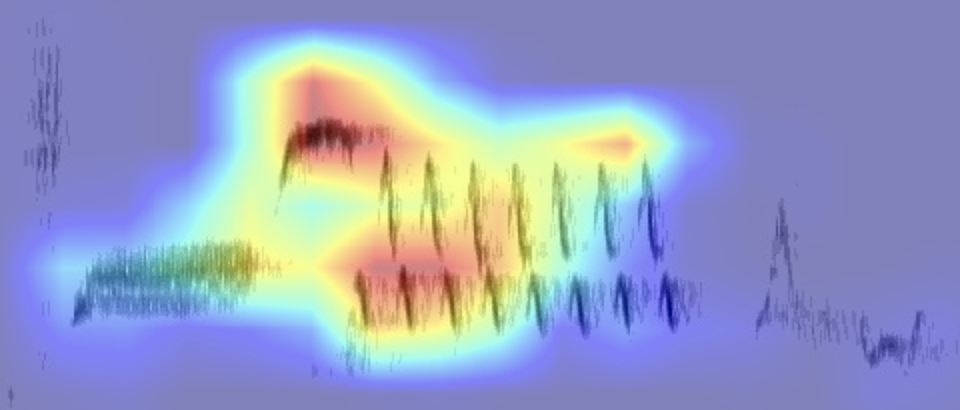}}} \hspace{0.0cm}
    \subfloat[Mexican, M2 sample DeepLIFT]{\fbox{\includegraphics[width=4.2cm, height=2cm]{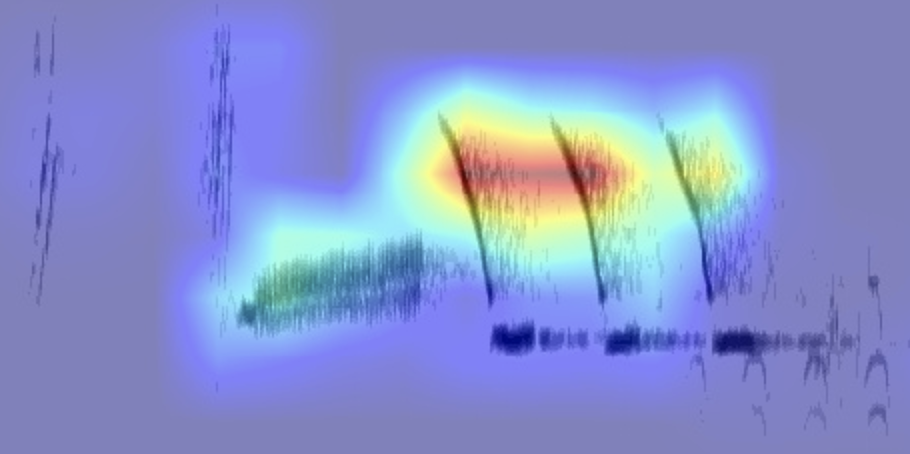}}}\hspace{0.0cm} 
    \subfloat[Mexican, M3 sample DeepLIFT]{\fbox{\includegraphics[width=4.2cm, height=2cm]{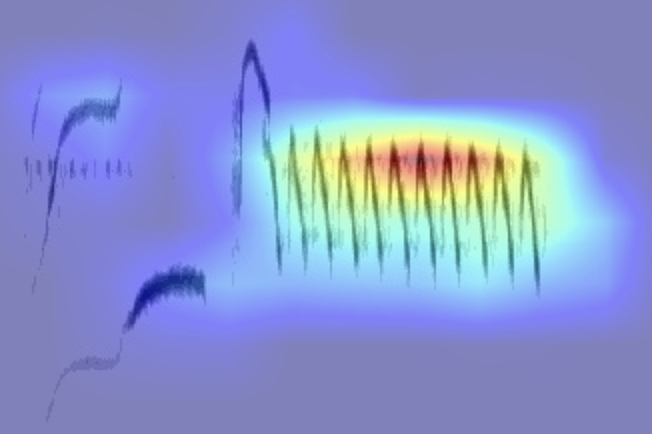}}}\hspace{0.0cm}
    \subfloat[Mexican, M4 sample DeepLIFT]{\fbox{\includegraphics[width=4.2cm, height=2cm]{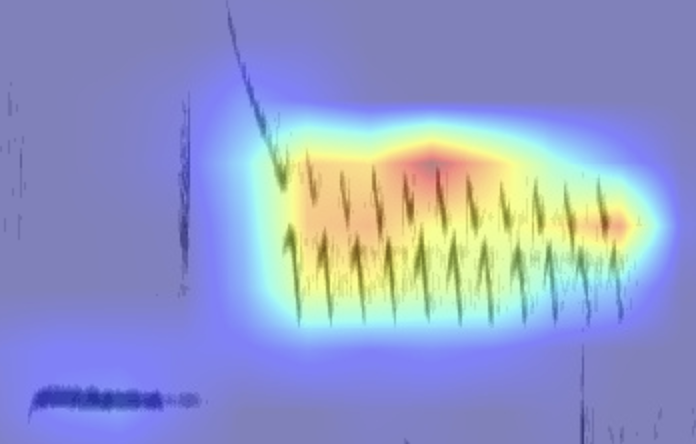}}}
    
    \caption{Figure shows the explanations obtained using Grad-CAM and DeepLIFT on samples selected from different t-SNE clusters. Fig (a) to (d) represent Grad-CAM explanations, and (e) to (h) represent DeepLIFT explanations from the Eastern variant. Fig (i) to (l) represent Grad-CAM explanations, and (m) to (p) represent DeepLIFT explanations from the Mexican variant. Visually, these samples look different, making the explanation less generalizable across the entire class.}
    \label{fig:clusters}
\end{figure*}

To generate more fine-grained explanations, Grad-CAM and DeepLIFT were applied. Fig~\ref{fig:eastern_mixed_Grad}(a) and (e) show heatmaps obtained using Grad-CAM for the Eastern and Mexican classes, respectively. These results show that Grad-CAM was able to characterize the entirety of spectrogram images in terms of importance, where red-colored segments are of high importance and blue-colored segments have low or zero importance in generating predictions. Similar experiments were performed using DeepLIFT and the results are shown in Fig~\ref{fig:eastern_mixed_Grad}(b) and (f).

For the Eastern class, both Grad-CAM and DeepLIFT consistently highlight low- to mid-frequency elements, particularly toward the end of the song. These salient elements also tend to have high amplitude. For the Mexican class, Grad-CAM and DeepLIFT often highlight parts of the trill at the end of the song. Grad-CAM often picks up reverberation following the end of the Mexican song, but this issue does not occur in the explanations of the Eastern class. DeepLIFT does not share this problem and highlights only the signal itself. DeepLIFT, therefore, produced the most interpretable explanations for bird song experts. Across both song types and methods, repeated elements near the end of songs are consistently highlighted as salient regions, as shown in Figures~\ref{fig:eastern_mixed_Grad}(a), (b), (e), and (f). These terminal trills sound different between the two variants and differ in spectro-temporal characteristics. They are often most distinctive to subjective human observers as well. These patterns not only appear across XAI methods but also align with how human observers distinguish between song types, suggesting that the model captures biologically relevant features.  These results show that XAI is promising for the future of bioacoustic research. Complex models can pick up on fine-grained variation in a complex biological dataset, and XAI can reveal that variation to human users. This allows bioacousticians and ecologists to fine-tune their acoustic classification models and can drive the generation of hypotheses for future study.

% Fig~\ref{fig:eastern_mixed_Grad}(c) and (d) shows ensemble explanation obtained using weighted average and element-wise max approach for the Eastern class. Similarly, ensemble explanatins are shown in Fig~\ref{fig:eastern_mixed_Grad}(g) and (h) for the Mexican class.

\subsection{Ensemble XAI Heatmap Explanation}
While both Grad-CAM and DeepLIFT techniques independently capture several core discriminative regions, they differ in their emphasis. 
% Grad-CAM tends to highlight more sharply localized regions, whereas DeepLIFT often attributes broader and more distributed importance. 
Notably, certain salient areas identified by one technique are either downplayed or entirely absent in the other. To effectively harness the complementary strengths of both approaches, we constructed ensemble saliency maps using two strategies: a linear average (Figure \ref{fig:eastern_mixed_Grad}(c), (g)) and an element-wise maximum (Figure \ref{fig:eastern_mixed_Grad}(d), (h)) of the normalized heatmaps. The max-based ensemble, in particular, ensures that the most strongly activated regions from either technique are preserved.

Results show that both ensemble approaches successfully capture all key activation regions identified by either method, ensuring that no class-relevant areas are omitted. This holistic view enhances the interpretability of the model’s decision-making process and provides stronger visual evidence to support classification outcomes. The qualitative improvement is especially apparent in cases where either Grad-CAM or DeepLIFT alone fails to highlight subtle but relevant patterns essential for the prediction. The weighted average approach tends to highlight the common saliency part from the Grad-CAM and DeepLIFT.

To quantify the density of strongly activated regions, we calculated the percentage of heatmap cells above varying importance thresholds (from 0.4 to 0.9). As shown in Figure~\ref{fig:threshold_curve}, the ensemble-max saliency map consistently activates a higher proportion of regions across thresholds compared to either Grad-CAM or DeepLIFT alone, suggesting its effectiveness in capturing both localized and distributed cues. The ensemble average also captures larger and more relevant important regions than a stand-alone technique.

\subsection{Sub-population Analysis}
During the experiments, we observed varying explanations for different samples within the same class, suggesting the possible existence of sub-classes (sub-groups) among the spectrogram samples of a given class. To examine the distribution of samples within each class, PCA~\cite{pca} and t-SNE~\cite{tsne} were employed to reduce the dimensionality of the data and to visualize their arrangement in the latent space, as illustrated in Figure~\ref{fig:pca_tsne}.
The PCA plot did not reveal much information, though separate regions could be seen for both classes.  All data points in the latent space were very tightly plotted, and no clusters/sub-groups within each class were found. However, the t-SNE plot revealed more distributed points of data scattered in a few different clusters for each class, which meant that even within the same class of signals, there were several sub-classes. This information is significant, as earlier experimental analyses indicated that the explanations were not generalizable, possibly due to the presence of sub-populations within each class.

Figure~\ref{fig:clusters} presents the explanation heatmaps generated by Grad-CAM and DeepLIFT for distinct clusters within the Eastern and Mexican bird song classes. The first row displays Grad-CAM heatmaps corresponding to four representative clusters of the Eastern class, while the second row shows the corresponding DeepLIFT explanations for the same samples. Similarly, the third and fourth rows depict Grad-CAM and DeepLIFT heatmaps, respectively, for four clusters of the Mexican class. The original spectrograms from different clusters exhibit noticeable visual differences, which were also confirmed through auditory inspection of the corresponding audio recordings. Notably, the explanation heatmaps remained consistent within each cluster, indicating that both Grad-CAM and DeepLIFT successfully captured cluster-specific patterns. These findings suggest that, although the songs were recorded within similar geographical regions, they exhibit distinct acoustic characteristics across sub-populations.

\section{Conclusion}\label{conclusion}

This work presents a CNN-based deep model for bioacoustics classification and various XAI techniques to uncover patterns in real-world bioacoustics data. First, data was collected from two different bird song variants, and then a CNN-based deep model was trained with a binary classification accuracy of $94.8\%$. To explain model predictions and find hidden patterns in data, both model-agnostic (LIME and SHAP) and model-specific (DeepLIFT and Grad-CAM) explanation techniques were employed.
Results reveal that model-specific methods, particularly DeepLIFT, provide more consistent and interpretable outputs, effectively highlighting salient regions in spectrogram images. The paper proposed two ensemble XAI strategies and results show that both provided better interpretation of model's decision making compared to standalone techniques.
% Grad-CAM consistently outperformed other methods in identifying relevant spectrogram regions, offering deeper insight into the model’s decision-making process. 
% SHAP was also able to identify a few relevant regions of the image, indicating that ensemble XAI methods need to be investigated. 
Additionally, variation within classes, reflected by distinct clusters in t-SNE plots, suggests that different song types form meaningful subgroups. Overall, DeepLIFT and Grad-CAM emerges as a powerful tool for enhancing interpretability in bird sound classification and for generating biologically relevant, testable hypotheses. 
In future work, we plan to focus on developing robust AI systems capable of generalizing across diverse acoustic environments and recording conditions.

\bibliographystyle{IEEEtran}
% \bibliography{IEEEabrv,main}
\bibliography{main}

\end{document}